\pdfoutput=1
\documentclass[journal,12pt,draftcls,onecolumn]{IEEEtran}
\IEEEoverridecommandlockouts

\usepackage[numbers]{natbib}

\usepackage[utf8]{inputenc}
\usepackage[pdftex]{graphicx}
\usepackage{amssymb, amsmath,amsthm, amsfonts}
\usepackage{tikz}
\usepackage{pgfplots}
\pgfplotsset{compat=1.17}
\usepgfplotslibrary{groupplots} 
\usetikzlibrary{pgfplots.groupplots}
\usepackage{thmtools,thm-restate}

\usepgfplotslibrary{fillbetween}
\usepackage{verbatim}
\usepackage{pifont}
\usetikzlibrary{patterns}
\title{Coded Shotgun Sequencing}
\author{Aditya Narayan}
\newcommand*\Y{\mathcal{Y}}
\usepackage[ruled,vlined,noend]{algorithm2e} 
\usepackage{float}

\usepackage{ulem}

\newtheorem{theorem}{Theorem}
\newtheorem{cor}[theorem]{Corollary}
\newtheorem{lemma}{Lemma}

\newtheorem{defn}{Definition}

\newcommand{\aln}[1]{\begin{align*}#1\end{align*}}
\newcommand{\al}[1]{\begin{align}#1\end{align}}

\newcommand{\1}{{\mathbf 1}}
\newcommand{\X}{{\mathcal X}}
\newcommand{\Z}{{\mathcal Z}}
\newcommand{\R}{{\mathbb R}}

\newcommand{\cov}{{\rm F}}
\newcommand{\recost}{{\rm A}}

\newcommand{\Co}{\cov}
\newcommand{\RC}{\recost}

\renewcommand{\emph}[1]{{\textit{#1}}}

\newif\ifdraft
\drafttrue

\newcommand{\Geometric}{{\rm Geometric}}

\newcommand{\TPC}{{\rm TPC}}

\begin{document}

\title{Recovering a Message from an Incomplete Set of Noisy Fragments\footnote{This paper was presented in part at the IEEE International Symposium of Information Theory 2021}}
\author{
Aditya~Narayan~Ravi, Alireza~Vahid, Ilan~Shomorony
\thanks{Aditya Narayan Ravi and Ilan Shomorony are with the Electrical and Computer Engineering Department of the University of Illinois, Urbana-Champaign, IL, USA. Email: {\sffamily anravi2@illinois.edu,ilans@illinois.edu}.}
\thanks{Alireza Vahid is with the Department of Electrical and Microelectronic Engineering at Rochester Institute of Technology, Rochester, NY, 14623, USA. Email: {\sffamily alireza.vahid@rit.edu}.}
}


\maketitle


\begin{abstract}
We consider the problem of 
communicating over a channel that breaks the message block into fragments of random lengths, shuffles them out of order, and deletes a random fraction of the fragments.
Such a channel is motivated by applications in molecular data storage and forensics, and we refer to it as the torn-paper channel.
We characterize the capacity of this channel 
under arbitrary fragment 
length distributions and deletion probabilities.
Precisely, we show that
the capacity is given by a closed-form expression that can be interpreted as 
$\Co - \RC$, where $\Co$ is the coverage fraction, i.e., the fraction of the input codeword that is covered by output fragments, and $\RC$ is an alignment cost incurred due to the lack of ordering in the output fragments.
We then consider a noisy version of the problem, where the fragments are  corrupted by binary symmetric noise. 
We derive upper and lower bounds to the capacity, both of which can be seen as $\Co - \RC$ expressions.
These bounds match for specific choices of  fragment length distributions, and they are approximately tight in cases where there are not too many short fragments.
\end{abstract}



\section{Introduction}
\label{Section:Introduction}


Consider the problem of transmitting a message by encoding it in a codeword, which is then torn up into fragments of random sizes. 
The codeword may be also subject to noise, and some of the fragments may be lost in the process.
The remaining fragments are received \emph{out of order} by a decoder who wishes to recover the original message. 
How does one optimally encode a message to protect against the tearing, the fragment losses, the noise, and the shuffling? How do the achievable rates depend on specific fragment length distributions, noise, and fragment deletion probabilities?

\begin{figure}[htb!]
\centering
\includegraphics[width=0.5\linewidth]{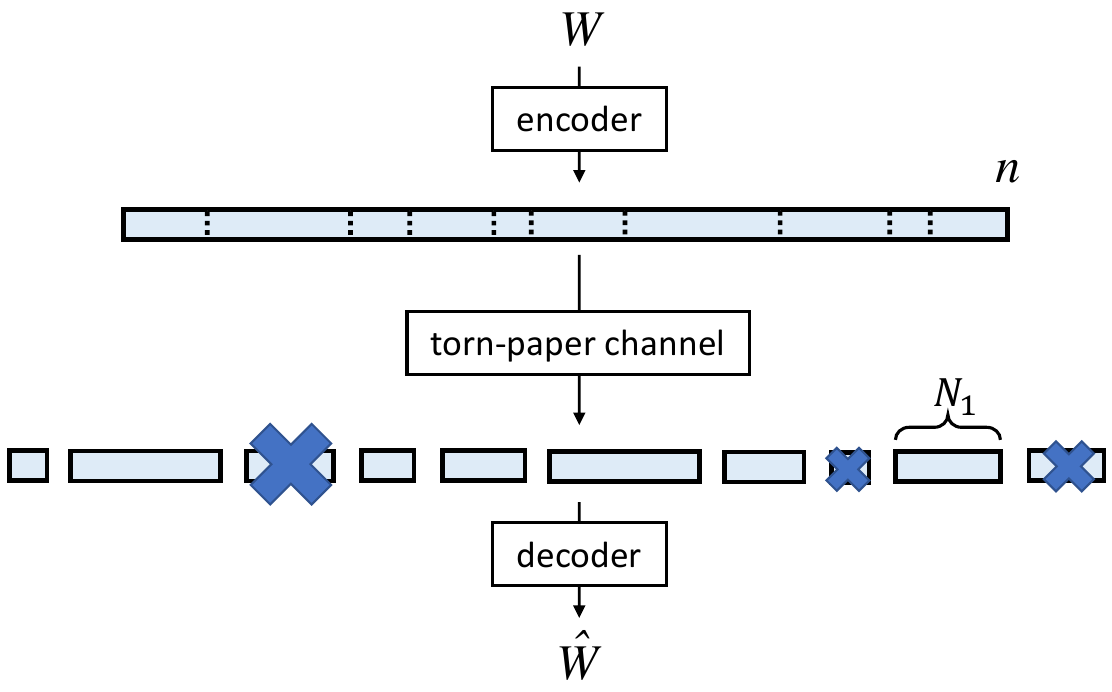}
\caption{The torn-paper channel with lost pieces. 
}
\label{fig:TPCLP}
\end{figure}

This channel was originally proposed in~\cite{TPCglobecom} as the \emph{torn-paper channel} (\TPC).
This channel is motivated by macromolecular data storage, and in particular DNA-based data storage~\cite{church_next-generation_2012,goldman_towards_2013,grass_robust_2015,bornholt_dna-based_2016,erlich_dna_2016,organick_scaling_2017}, where the data is encoded into molecules that may be subject to breaks during storage.
Moreover, when retrieving the data via sequencing, molecules are read in a random order, and many fragments are lost
\cite{heckel_characterization_2018}.
It can also be motivated by applications in fingerprinting and forensics, where one may wish to encode a serial number into a physical object (such as a weapon), which should be recoverable even from a small set of pieces left over from the original object \cite{forensicsfragment,wang2023breakresilient}. 
In this paper, we generalize the torn-paper channel to 
capture several 
real-world challenges of DNA storage. Precisely, we consider a \TPC~whose
 input is a length-$n$ binary string, which is broken up into $K$ fragments of random lengths $N_1,\dots,N_K$, such that $N_1 + \dots + N_K  =n$,
 where the number of fragments $K$ is itself random.
Each resulting fragment can be independently deleted with a probability $d$, which in general is allowed 
to be a function $d(\cdot)$ of the fragment length $N_i$.
The channel output is an unordered multiset of all the fragments that are not deleted. 
This torn-paper channel with lost pieces
is illustrated in Figure~\ref{fig:TPCLP}. 

The original TPC introduced in~\cite{TPCglobecom} is a special case of this channel  where there are no fragment deletions ($d = 0$) and the fragment lengths are distributed as $N_i \sim \Geometric(1/\ell_n)$ (where $E[N_i] = \ell_n$ can be a function of $n$).
The capacity in this setting was shown in~\cite{TPCglobecom} to be
\al{
C_{\rm TPC} = \exp \left( {-\lim_{n\to \infty}  \frac{\log n}{\ell_n}}\right).
\label{eq:ctpc}
}
By defining $\alpha \triangleq \lim_{n\to \infty} (\log n)/\ell_n$, we can also write $C_{\rm TPC} = e^{-\alpha}$.
The simplicity and ``elegance'' of the capacity expression $C_{\rm TPC} = e^{-\alpha}$ is a result of the specific choice of geometrically-distributed fragment lengths and the connection between the geometric and exponential distributions.
As such, it is unclear how \eqref{eq:ctpc} would generalize to different fragment distributions and to the case of lost fragments.

Another related result we consider that can help build intuition
is the capacity of the 
\emph{shuffling channel} \cite{noisyshuffling}. 
The input to the shuffling channel is a multiset 
of strings of a \emph{fixed length}, which are shuffled by the channel.
The shuffling channel can be thought of as a TPC where the fragment lengths are all \emph{deterministic} and equal; i.e., $N_1 = \dots = N_K = \ell_n$, and $K = n/\ell_n$.
Notice that, in this case, the encoder knows all the tearing locations (since they are deterministic) and can combat the lack of ordering at the output by placing a unique index at the beginning of each fragment.
The results in \cite{noisyshuffling} show that the capacity of this shuffling channel is
\al{
C_{\rm shuf} = \left( 1- \lim_{n\to \infty} \frac{\log n}{\ell_n} \right)^+ = (1-\alpha)^+,
\label{eq:cshuf}
}
where $(x)^+ \triangleq \max(0,x)$.
As explained in \cite{noisyshuffling}, the term $(\log n)/\ell_n$ can be understood as the fraction of bits in each length-$\ell_n$ fragment that must be used for a unique index, which allows for the alignment of the shuffled fragments.
We can thus think of $\alpha = \lim_{n\to \infty} \frac{\log n}{\ell_n}$ as a fundamental alignment cost.
Notice that the results in \eqref{eq:ctpc} and \eqref{eq:cshuf} feel qualitatively different, and it is not clear how they can both be seen as special cases of a general TPC capacity.




The first main result in this paper generalizes the capacity of the TPC to (i) accommodate the case of lost fragments with a general deletion probability function $d(\cdot)$, and (ii) allow any distribution for the fragment length $N_i$, as long as some mild regularity conditions hold. 
In Section~\ref{SubSection:TPCLP-MainResults}, we obtain closed-form expressions for various choices of $d(\cdot)$ and $N_i$.
Moreover, in doing so, we provide a capacity expression that allows us to reconcile \eqref{eq:ctpc} and \eqref{eq:cshuf}.
 More precisely, we prove that discarding fragments of length $\log n$ or shorter at the output does not affect the capacity, and that the capacity of the TPC 
 can always be written as
\al{
C_{\TPC} = \cov_d{\{\log{n}\}} - \recost_d{\{\log{n}\}}
\label{eq:cov-recost}.
}
Here, $\cov_d\{\log{n}\}$ (the covered fraction) represents the fraction of the original length-$n$ string that is covered 
by the non-deleted output fragments (with deletion probability $d$) and after discarding those shorter than $\log n$. 
The alignment cost $\recost_d\{\log{n}\}$ represents the fraction of the non-deleted output fragments that would need to be dedicated for the placing of a unique index to help align the fragments if we knew the tearing locations (again after discarding fragments shorter than $\log{n}$)\footnote{The terms $F$ and $A$ are referred to as ``coverage''  ($\Phi$) and ``reordering cost'' ($\Omega$) in the conference version of this paper \cite{TPCLP}. Here, we adopt $F$ and $A$ in an attempt to improve the notation.}. Surprisingly, it turns out that this alignment cost does not change in the setting of the TPC 
where the tearing points are unknown a priori even when 
the encoder cannot place unique indices at the beginning of each fragment.

Consider the capacity expression for the deterministic case with fragments of length $\ell_n > \log{n}$ and no lost fragments. 
Notice that the covered fraction for this case is $1$, since all the fragments are retained. 
Now, the alignment cost is the fraction of bits in each length-$\ell_n$ piece used for indexing purposes given by $\lim_{n \to \infty} \log n/\ell_n = \alpha$ as discussed before. 
Thus, the formula ``\Co $-$ \RC'' recovers (\ref{eq:cshuf}) as a special case. 
Similarly, the result in \cite{TPCglobecom} for a TPC with geometric piece lengths and no lost fragments
is a special case of (\ref{eq:cov-recost}).
More specifically,
for the setting in~\cite{TPCglobecom} 
the covered fraction can be calculated as $(1 + \alpha)e^{-\alpha}$ and the alignment cost can be shown to be  $\alpha e^{-\alpha}$, which yields (\ref{eq:ctpc}) (see \cite{TPC-journal} for details).

The general expression in \eqref{eq:cov-recost} captures in an intuitive way the impact of sequence fragmentation and fragment losses, but it is based on a model where the fragments are observed in an error-free fashion at the output. 
In many applications, including macromolecular data storage, one can observe bit-level substitution errors, or bit flips. 
Channels such as the shuffling channel have been extended to accommodate these kinds of errors \cite{noisyshuffling, lenz_upper_2019}. 
Notably, \cite{noisyshuffling} considers the capacity of the noisy shuffling channel, where the input string is passed through a Binary Symmetric Channel (BSC) with crossover probability $p$. 
Specifically, the results in \cite{noisyshuffling} show that if the fragment size is 
$\ell_n \geq \frac{2 \log n}{1 - H(2p)}$, then the capacity is

\begin{align} \label{eq:noisyshuffcap}
    C_{\text{noisy-shuf}} = 1 - H(p)  - \alpha.
\end{align}

\begin{figure}[htb!]
\centering
\includegraphics[width=0.5\linewidth]{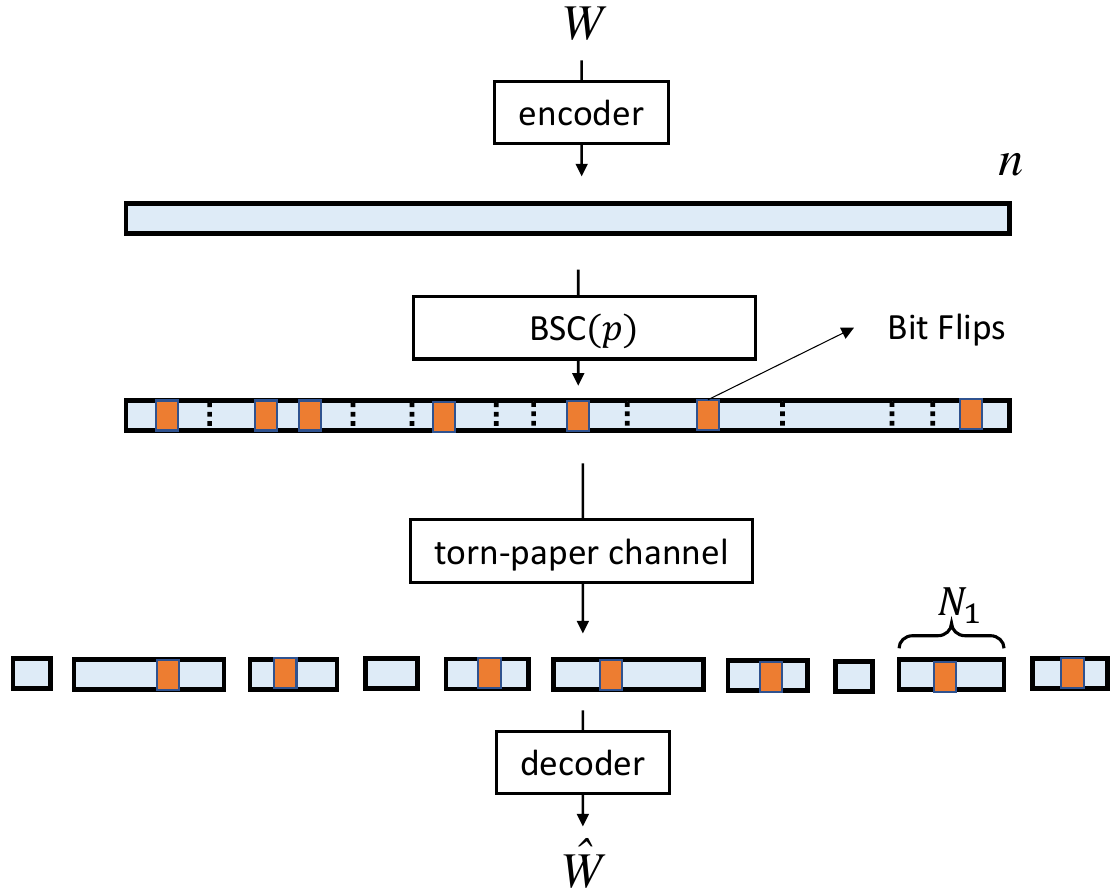}
\caption{The noisy torn-paper channel. 
}
\label{fig:TPCNoisy}
\end{figure}

Therefore to study the impact of symbol-level noise on the TPC capacity, we also study a TPC where the input 
is passed symbol-wise 
through a Binary Symmetric Channel (BSC) with crossover probability $p$. 
This is followed by a tearing of the resulting string in the exact same fashion as before.
To keep things simple, we assume no fragments are lost, but the results can be easily extended to accommodate lost fragments too.
The Noisy-TPC is shown in Figure~\ref{fig:TPCNoisy}.

The second main result in this paper provides inner and outer bounds to the capacity $C_{\text{Noisy-TPC}}$ for arbitrary distributions of fragment length $N_i$ and crossover probability $p$. 
Following the intuition from the $\cov{} - \recost{}$ expression, a natural conjecture is that the capacity $C_{\text{noisy-TPC}}$ in this case is given by
\begin{align} \label{eq:conj}
   xc = C_{\rm BSC} \cdot \cov{\left\{\frac{\log{n}}{C_{\rm BSC}}\right\}} -\recost{\left\{\frac{\log{n}}{C_{\rm BSC}}\right\}},
\end{align}
where  $C_{\rm BSC} = 1 - H(p)$ is 
the capacity of a BSC($p$) channel.
Here, $\cov$  and $\recost$ 
have the same meaning as before, except only pieces of lengths greater than the lengths indicated in the brackets are considered.
Since the fragments are now noisy, 
we expect the covered fraction F
to be multiplied by $C_{\rm BSC}$,
since that is effectively the fraction of ``good'' bits recovered. 
Moreover, due to the noise, we expect the ``effective'' length of a given fragment  (call it $N_{\rm eff}$) to be related to its original length $N$ as
$
    N_{\rm eff} = C_{\rm BSC} \times N.
$
Since we discarded all strings of size less than $\log{n}$ for the noiseless case, it is natural to conjecture that we should discard all pieces with $N_{\rm eff} \leq \log{n}$ while calculating coverage fraction and alignment cost, which is equivalent to discarding fragments with $N \leq \log{n}/C_{\rm BSC}$.

As it turns out, we can indeed construct codes that achieve the rate in (\ref{eq:conj}). 
More precisely we prove that 
rate $R$
is achievable if
\begin{align}\label{eq:NoisyInnerBound}
    R < (1-H(p))\cov{\left\{\frac{\log{n}}{1-H(p)}\right\}} -\recost{\left\{\frac{\log{n}}{1-H(p)}\right\}}.
\end{align}
However, deriving an outer bound that matches this inner bound is hard in general.
Instead, we prove that no rates $R$ with 
\begin{align}\label{eq:NoisyOuterBound}
    R > (1-H(p))\cov{\left\{\log{n}\right\}} -\recost{\left\{\frac{2\log{n}}{1-H(2p)}\right\}}
\end{align}
are achievable.
Notice that the difference between (\ref{eq:NoisyInnerBound}) and (\ref{eq:NoisyOuterBound}) is in the length of the pieces we discard when computing \Co~and \RC.
As such, as long as the fragment lengths $N_i$ are all guaranteed to be large enough, the inner and outer bounds match.

Specifically, we can show that for the class of noisy TPCs where \(N_1 \geq \frac{2\log{n}}{1-H(2p)}\) with probability \(1\) (i.e., every fragment is longer than this threshold), the capacity of the noisy TPC is given by
\begin{align}
    C_{\text{noisy-TPC}} = 1 - H(p) - \alpha.
\end{align}
Interestingly, this result generalizes the capacity of the noisy shuffling channel when \(\ell_n \geq \frac{2\log{n}}{1-H(2p)}\). 
Therefore, we can view the result for the noisy shuffling channel in \eqref{eq:noisyshuffcap} as a special case of this more general result that allows fragments to have random lengths.

\noindent \textbf{Related Work:}
The torn-paper channel we consider is a generalization of the channel 
from \cite{TPCglobecom, TPC-journal, TPCLP}, originally introduced with the goal of modeling message fragmentation in  macromolecular data storage and especially DNA-based data storage. 
As such, our work is related to other attempts at modeling the DNA storage channel and characterizing its capacity.
This includes the noisy shuffling channel model~\cite{DNAStorageISIT,noisyshuffling,DNAStorageIT} and multi-draw sampling channels~\cite{lenz_upper_2019,lenz2020achieving}.
See~\cite{DNAStorageMonograph, tmbmcoutoforder} for a broader survey of the topic.

The idea of index-based coding was used in \cite{bar2022adversarial} to develop an encoding-decoding scheme for an adverserial version of the TPC. Specifically it considered a TPC where the length of the torn pieces were restricted between values $L_{\rm min}$ and $L_{\rm max}$ with probability $1$. Moreover \cite{bar2022adversarial} proves that decoding can be done in linear time with this coding scheme.


The torn-paper channel we consider is also related to the shotgun sequencing channel~\cite{ravi2021coded, isitshotgun}, whose output also consists of pieces of the input codewords.
However, the shotgun sequencing channel models standard next-generation sequencing platforms, where the observed output fragments have overlaps with each other~\cite{MotahariDNA,gabrys2018unique}.
Moreover, an ``extreme'' version of a torn-paper channel that breaks the message down to its symbols (i.e., fragments of length one) and shuffles them was studied as the noisy permutation channel~\cite{makur2018information,TangPolyanskiyNoisyPerm}.


In addition to information-theoretic characterizations of the fundamental limits of DNA storage channels, several recent works have proposed new explicit code constructions based on unique aspects of DNA data storage. 
These works focus on DNA synthesis constraints such as sequence composition  \cite{kiah_codes_2016,yazdi_rewritable_2015,erlich_dna_2016}, the asymmetric nature of the DNA sequencing error channel \cite{gabrys_asymmetric_2015}, 
the need for codes that correct insertion errors \cite{sala_insertions_2016}, 
and techniques to allow random access \cite{yazdi_rewritable_2015}. 
 Moreover \cite{embeddednon-overlap, embeded-merge} propose various VT codes for the TPC model considered in \cite{TPC-journal}.








\section{The Torn-Paper Channel with Lost Pieces}

In this section, we consider a version of the TPC in which some fragments are lost. 
The converse involves a careful decomposition of the TPC into parallel channels that contain pieces of roughly the same lengths. The converse is inspired by the converse proved in \cite{TPCglobecom}, but requires considerable generalization to account for general distributions of $N_i$ and lost fragments.
We present an achievable argument based on random coding. 
The resulting inner and outer bounds match, characterizing the capacity of this channel.
We will consider the noisy TPC in Section~\ref{Section:TPCLPNoisy}.


\label{Section:TPCLP}
\subsection{Problem Setting}
\label{SubSection:TPCLP-ProbSetting}
We consider the \TPC{} as shown in Figure~\ref{fig:TPCLP}.
The transmitter encodes a binary codeword $X^n \in \{0,1\}^n$ corresponding to the message $W \in [1:2^{nR}]$. 
The channel output is a multiset of variable-length binary strings $\Y$. The process by which $\Y$ is obtained from $X^n$ is described as follows.
 The channel first breaks the input sequence into pieces of a random length.
 Specifically, define $N_1,N_2,\dots$ to be i.i.d.~random variables. 
 We assume $E[N_i] = \ell_n$ $\forall$ $i$. Let $K$ be the smallest index such that $\sum_{i=1}^{K}N_i$ is greater than or equal to $n$. Note that $K$ is also a random variable. The channel tears the string $X^n$ into $\vec{X}_1,\vec{X}_2,\dots,\vec{X}_K$ where
    \begin{align}
        & \vec{X}_i \triangleq \left[X_{1 + \sum_{j=1}^{i-1}N_j},\dots,X_{\sum_{j=1}^{i}N_j}\right] \text{ and} \nonumber \\
        & \vec{X}_K \triangleq \left[X_{1 + \sum_{j=1}^{K-1}N_j},\dots,X_n\right]. \nonumber
    \end{align}
    
    Then, the multiset 
 $\Y$ (which is the output of the channel) is defined as follows. 
 Consider the strings $\{\vec{X}_1,\vec{X}_2,\dots,\vec{X}_K\}$. Each element $\vec{X}_i$ is independently discarded with probability $d(N_i)$, where $d(\cdot)$ is the deletion probability (which in general depends on the length of the pieces), for a function $d : \{1,2,\dots\} \to [0,1]$. 
 The new multi-set obtained is $\Y$, which is the output of the channel. 
Note that there are no bit-level errors (e.g., deletions or bit-flips). Moreover, $\ell_n$ in general depends on the value of $n$.

For the purposes of this paper we assume (i) the limit $\alpha \triangleq \lim_{n \to \infty} \log n/\ell_n$ exists and $\alpha \in (0,\infty)$
and (ii) $E[N_1^2/(\log{n})^2]$ is finite and bounded for all $n$. 
This means we expect the second moment of  $N_i$ to scale as $\log^2{n}$. This is typically valid when pieces are of a size that scales as $\log{n}$ when we increase $n$.
Our results also hold in the case where $\alpha = 0,\infty$, but would require different steps to prove. 
%



\noindent \textbf{Notation:} Throughout the paper, $\log(\cdot)$ represents the logarithm in base $2$. For functions $a(n)$ and $b(n)$, we say $a(n) = o(b(n))$ if $a(n)/b(n) \to 0$ as $n \to \infty$. 
For an event $A$, we let $\mathbf{1}_A$ be the binary indicator of $A$. 
For $x \in \R$, 
$\lceil x \rceil$ 
refers to the smallest integer greater than equal to $x$.
\subsection{Capacity of the generalized TPC}
\label{SubSection:TPCLP-MainResults}

Intuitively, the capacity of the TPC should be affected by 
two distinct sources of uncertainty:
(i) some of the fragments (which potentially carry information) are discarded by the channel, and (ii) the 
remaining fragments are observed as an unordered set.
As it turns out, (i) will be captured in the capacity expression by a quantity 
that represents the fraction of bits in $X^n$ retained at the output, which can be written as
\begin{align}
     \frac{1}{n}\sum_{i=1}^{K}N_i\mathbf{1}_{\{\vec{X}_i \in \Y\}}.
     \label{eq:cov0}
\end{align}
The limit as $n \to \infty$ of the expected value of (\ref{eq:cov0}), after ignoring fragments smaller than a fixed number $g(n)$, turns out to be fundamental in calculating the capacity of the \TPC{}. 


\begin{defn}\label{defn:Coverage}
The coverage fraction $\cov_{d}\{g(n)\}$ ($d$ is suppressed if $d \equiv 0$) is defined as 
\begin{align}
    \cov_{d}\{g(n)\} \triangleq \lim_{n \to \infty}E\left[\frac{1}{n}\sum_{i=1}^{K}N_i\mathbf{1}_{\left\{\vec{X}_i \in \Y, N_i \geq g(n)\right\}}\right].
\end{align}
\end{defn}
The capacity of the \TPC{} will also involve a quantity that captures (ii).
To build intuition, let us imagine a channel where the tearing points and the set of pieces that are discarded are known a priori. 
For this channel, to preserve the ordering, a simple coding strategy is to include an index/address at the beginning of each fragment. There are roughly $n/\ell_n$ fragments prior to some of them being lost.
We therefore need
\begin{align*}
    \log{(n/\ell_n)} &= \log{n} - \log{\ell_n} \stackrel{(a)}{=} \log{n} - o(\log{n})
\end{align*}
bits per piece for indexing, where (a) holds since $\alpha \in (0,\infty)$ and $1/\ell_n = \alpha/\log{n}$ asymptotically.
Therefore, for a piece $\vec{X}_i$ in $\Y$, $\log{n}$ bits are needed for indexing. 
This can be written as $\mathbf{1}_{\{\vec{X}_i \in \Y\}}\log{n}$ and can be thought of as an ``alignment cost'' that one must pay to know where in a codeword $\vec{X}_i$ belongs.  
Surprisingly, this cost remains unchanged even when the tearing locations are unknown.
The 
average 
of this quantity across all pieces also 
plays a key role in
the capacity of the \TPC.
\begin{defn}\label{defn:ReorderingLoss}
    The alignment cost $\recost_{d}\{g(n)\}$ ($d$ is suppressed if $d \equiv 0$) is defined as
    \begin{align}
       \recost_{d}\{g(n)\} \triangleq \lim_{n \to \infty}E\left[\frac{\log{n}}{n}\sum_{i=1}^{K}\mathbf{1}_{\left\{\vec{X}_i \in \Y,N_i\geq g(n)\right\}}\right].
    \end{align}
\end{defn}

We now state our main result.
\begin{theorem}\label{thm:Main}
The capacity of the \TPC{} is 
\begin{align}
    C = \cov_d\{\log{n}\} - \RC_d\{\log{n}\}.
\end{align}
\end{theorem}
The capacity expression is of the form ``$\Co - \RC$'', which intuitively is the fraction of bits that carry information about the message.
 The following corollary allows us to compute the capacity numerically. It is also used to obtain closed form expressions for various distributions of $N_i$.
\begin{restatable}{cor}{thmMain}
\label{thm:Main2}
Assuming limits exist, the capacity of the \TPC{} is equivalently written as
\begin{align}\label{eq:cap_intro}
    C = \alpha\int_{1}^{\infty}\left(\beta - 1\right)\left(1 - \hat{d}(\beta)\right)h(\beta)d\beta,
\end{align}
where $\hat{d}(\beta) \triangleq \lim_{n \to \infty} d(\beta \log{n})$ and $h(\beta) \triangleq \lim_{n \to \infty}\Pr(N_1 = \beta\log{n})\log{n}$.
\end{restatable}
We prove Corollary~\ref{thm:Main2} in Appendix~\ref{app:cor2}. 
Note that the result in \cite{TPCglobecom} for the TPC with no lost pieces can be obtained by taking $\hat{d}(\beta) = 0$. 
To see this note that when $N_1 \sim \rm Geom(1/\ell_n)$, we can compute $h(\beta)$ as
\begin{align}\label{eq:geomh_beta}
h(\beta) = \lim_{n \to \infty} \left(1-\frac{1}{\ell_n}\right)^{\beta\log{n}}\frac{\log{n}}{\ell_n} =  \alpha e^{-\alpha\beta}.
\end{align}
Using this to compute \eqref{eq:cap_intro} we obtain
\begin{align}
    C = \alpha\int_{1}^{\infty}(\beta - 1)(\alpha e^{-\alpha\beta})d\beta = \left(1 + \alpha\right)e^{-\alpha} - \alpha e^{-\alpha} = e^{-\alpha},
\end{align}
which is the capacity result discussed in Section~\ref{Section:Introduction} and first proved in \cite{TPCglobecom}.
Similarly, we can find a closed-form expression for the capacity of several different choices of $N_i$ and the deletion probability $d(\cdot)$, as shown in
Table~\ref{tab:table1}.
Note that $h(\beta)$ can have multiple $d(\cdot)$s associated with it.

\begin{table}[ht]
\vspace{-4mm}
  \begin{center}
    \caption{Capacity Expressions    \label{tab:table1}}
    \vspace{-1mm}
    \begin{tabular}{c|c|c|c} 
      $\hat{d}(\beta)$ & $N_i$ & $h(\beta)$ & $C_{\text{\TPC}}$\\
      \hline
      $0$ & \Geometric($1/\ell_n$) & $\alpha e^{-\alpha\beta}$ & $e^{-\alpha}$\\
      $\epsilon$ & \Geometric($1/\ell_n$) & $\alpha e^{-\alpha\beta}$ & $(1- \epsilon)e^{-\alpha}$\\
       $e^{-\gamma\beta}$ & \Geometric($1/\ell_n$) & $\alpha e^{-\alpha\beta}$ & $e^{-\alpha}\left(1 - \frac{\alpha^2e^{-\gamma}}{(\alpha + \gamma)^2}\right)$\\
      $0$ & $\text{U}[0:\gamma\log{n}]$, $\gamma \geq 1$ & $1/\gamma$ &$\left((\gamma - 1)/\gamma\right)^2$ \\
       $0$ & \text{Fixed}($\ell_n$), $\ell_n \geq \log{n}$ & NA\footnote{$h(\cdot)$ does not exist, hence we directly employ Theorem~\ref{thm:Main}.} & $1-\alpha$
      \vspace{3mm}
    \end{tabular}
    
    \footnotesize{$^2$ $h(\cdot)$ does not exist, hence we directly employ Theorem~\ref{thm:Main}.}
  \end{center}
  \vspace{-3mm}
\end{table}
\subsection{Achievability via Random Coding}
\label{SubSection:TPCLP-AchievableRates}

We use a random coding argument to prove the achievability of Theorem~\ref{thm:Main}. 
We generate a codebook $\mathcal{C}$ with $2^{nR}$ codewords, by independently picking each letter as $\text{Bern}(1/2)$. 
Let the resulting random codebook be $\mathcal{C} = \{\mathbf{x}_1,\mathbf{x}_2,\dots,\mathbf{x}_{2^{nR}}\}$. 

Assume that $W=1$ is the message that is transmitted. 
The output $\Y$ is available at the decoder. 
We follow steps similar to \cite{TPCglobecom}, but with considerable generalization. 
We choose a sub-optimal decoder, which throws out all fragments of size at most $\log{n}$. 
Let this set be $\Y_{\geq 1}$.
If elements of $\Y_{\geq 1}$ exist as non-overlapping substrings in a single codeword $\mathbf{x}_i$, then the decoder declares the index of that codeword as the message. 
Since there is no noise in the observed fragments, the only error event corresponds to the existence of an incorrect codeword containing all fragments in $\Y_{\geq 1}$.
We bound the probability of error averaged over all codebook choices as
\begin{align}
    &\text{Pr}(\mathcal{E}) = \text{Pr}(\mathcal{E}|W = 1) \nonumber\\
     &= \text{Pr}\left(\exists \text{ } j\neq1:x_j\text{ contains all fragments in $\Y_{\geq 1}$}\middle|W = 1\right).
\end{align}

We now state two lemmas (the proofs of which are available in Appendices~\ref{app:ConcBoundCov}~and~\ref{app:RCLem}) that are crucial to prove the achievable part of Theorem~\ref{thm:Main}. They provide us with a concentration on the coverage fraction and the alignment cost. This intuitively can be used to bound the probability of error by restricting the possibility of the value of the coverage fraction $\Co$ being too high or the alignment cost $\RC$ being too low, since they are events of low probability. Let us define the event $\tilde{B} := \{\vec{X}_1 \in \Y_{\geq 1}\}$. 
Then the following lemmas hold.
\begin{restatable}{lemma}{CoverageLemma}
\label{Lemma:ConcBoundCoverage}
For any $\epsilon > 0$, as $n \to \infty$, 
\begin{align}\label{eqn:ConcBoundCoverage}
    \Pr\left(\left|\frac{1}{n}\sum_{i=1}^{K}N_i\mathbf{1}_{\{\vec{X}_i \in \Y_{\geq 1}\}} - \frac{E[N_1\mathbf{1}_{\tilde B}]}{\ell_n}\right| >  \frac{E[N_1\mathbf{1}_{\tilde B}]}{\ell_n}\epsilon\right) \to 0.
\end{align}
 In particular, this implies that $\Co_d\{\log n\} = \lim_{n \to \infty} \frac{E[N_1\mathbf{1}_{\tilde B}]}{\ell_n}$.
\end{restatable}
\begin{restatable}{lemma}{ConcBound}\label{Lemma:ConcBoundReordering}
For any $\epsilon > 0$, as $n \to \infty$,
\begin{align}\label{eqn:ConcBoundReordering}
    \Pr\left(\middle|\sum_{i=1}^{K}\mathbf{1}_{\{\vec{X}_i \in \Y_{\geq 1}\}} - \frac{nE[\mathbf{1}_{\tilde B}]}{\ell_n}\middle|\geq \frac{nE[\mathbf{1}_{\tilde B}]}{\ell_n}\epsilon\right)
    \to 0.
\end{align}
In particular, this implies that $\RC_d\{\log n\} = \lim_{n \to \infty} \frac{\log(n) E[\mathbf{1}_{\tilde B}]}{\ell_n}= \alpha \lim_{n\to \infty} E[\mathbf{1}_{\tilde B}]$.
\end{restatable}
Now we let $B_1 = (1+\epsilon)\frac{nE[\mathbf{1}_{\tilde B}]}{\ell_n}$ and $B_2 = (1-\epsilon)\frac{E[N_1\mathbf{1}_{\tilde B}]}{\ell_n}$ and define the event
\begin{align}
    \mathcal{B} = \left\{\sum_{i=1}^{K}\mathbf{1}_{\{\vec{X}_i \in \Y_{\geq 1}\}} > B_1\right\}\cup\left\{\frac{1}{n}\sum_{i=1}^{K}N_i\mathbf{1}_{\{\vec{X}_i \in \Y_{\geq 1}\}} < B_2\right\}.
\end{align}
Lemmas~\ref{Lemma:ConcBoundCoverage} and \ref{Lemma:ConcBoundReordering} imply that $\text{Pr}(\mathcal{B}) \to 0$ as $n \to \infty$. Therefore, we have
\begin{align}\label{eq:ProbUpperBound}
    \text{Pr}(\mathcal{E}) &= \text{Pr}(\exists j\neq1: x_j\text{ containing all fragments in $\Y_{\geq 1}$}|W = 1) \nonumber \\
    &\stackrel{(a)}{\leq}\text{Pr}(\exists j\neq1 :x_j\text{ containing all fragments in $\Y_{\geq 1}$}|W = 1,\overline{\mathcal{B}}) \nonumber \\
    &+ \text{Pr}(\mathcal{B}) \nonumber \\
    &\stackrel{(b)}\leq |\mathcal{C}|\frac{n^{B_1}}{2^{nB_2}} + \text{Pr}(\mathcal{B})\nonumber \\
    &= 2^{nR}2^{B_1\log{n}}2^{-nB_2} + o(1).
\end{align}

Inequality $(a)$ follows from the law of total probability and the fact that $\mathcal{B}$ is independent of $W=1$. Inequality $(b)$ follows from the union bound and the fact that given $\overline{\mathcal{B}}$, there are at most $n^{B_1}$ ways to align the fragments in $\Y_{\geq 1}$ to a codeword $x_j$. 
To see this, note that, given $|\Y_{\geq 1}| \leq B_1$, there are at most $n$ places the fragments can start from to align each piece and at most $B_1$ such pieces. 
Since a non-overlapping alignment of the strings in $\Y_{\geq 1}$ to a codeword $x_j$ covers at least $n B_2$ positions of $x_j$, the probability that it matches $x_j$ when $j\ne 1$, on all covered positions is at most $2^{-nB_2}$.
Now, $\Pr(\mathcal{E}) \to 0 $ if
\begin{align}\label{eq:AchFinalSteps}
    R & < \lim_{n \to \infty}\left(B_2 -\frac{\log{n}}{n}B_1\right) \nonumber\\
    &= \lim_{n \to \infty} \left((1-\epsilon)\frac{E\left[N_1\mathbf{1}_{\tilde B}\right]}{\ell_n} - (1+\epsilon)\frac{E\left[\mathbf{1}_{\tilde B}\right]\log{n}}{\ell_n}\right) \nonumber \\
    &= (1-\epsilon)\cov_d\{\log{n}\} - (1+\epsilon)\RC_d\{\log{n}\}.
\end{align}
Letting $\epsilon \to 0$, we conclude that any rate 
$R < \cov_d\{\log{n}\} - \RC_d\{\log{n}\}$ is achievable.
This proves the achievability part of Theorem~\ref{thm:Main}.

\subsection{Converse}
\label{SubSection:TPCLP-Converse}
In order to prove the converse, we partition the set $\Y$ into sets that contain pieces of roughly the same length. 
This allows us to view the \TPC~ 
as a set of parallel channels that process pieces of roughly the same length. More precisely, we define
\begin{align}\label{eq:y_kdef}
    \Y_k \triangleq \left\{\vec{X}_i \in \Y: \frac{k-1}{L}\log{n} \leq N_{i} < \frac{k}{L}\log{n}\right\}, 
\end{align}
where $L$ is a fixed integer. We then split the set of ``channels'' into two sets, one with pieces of smaller sizes and the other with larger sizes. Specifically, we fix another integer $J > L$, and define $\Y_{\geq J} = \{\vec{X}_i:N_i \geq (J/L)\log{n}\}$.
Then, by Fano's inequality, we have
\begin{align}\label{eq:Fano}
    R &\leq \lim_{n \to \infty} \frac{I(X^n;\Y)}{n} \leq \lim_{n \to \infty} \frac{H(\Y)}{n} \stackrel{(a)}{\leq} \lim_{n \to \infty} \sum_{k = 1}^{J}\frac{H(\Y_k)}{n} + \lim_{n \to \infty}\frac{H(\Y_{\geq J})}{n},
\end{align}
where (a) holds from the independence bound on partition $\Y =(\cup_{k=1}^{J}\Y_k)\cup\Y_{\geq J}$. 


The key idea is that for fixed large values of $J$, the second term in equation \eqref{eq:Fano} is finite, but arbitrarily small. 
We will now use the fact that $|\Y_k|$ concentrates around its mean to tackle the first term in \eqref{eq:Fano}.
To that end, we define the event
$B = \{X_1 \in \Y\}$ 
and
\begin{align}
    q_{k,n} &= \text{Pr}\left(\frac{k-1}{L}\log{n} \leq N_1 < \frac{k}{L}\log{n}\right) \nonumber \\
    e_{k,n} &= \text{Pr}\left(B\,\middle|\,\frac{k-1}{L}\log{n} \leq N_1 < \frac{k}{L}\log{n}\right).
\end{align}
Additionally, we define the event 
\aln{
\mathcal{E}_{k,n} =
\{||\Y_k|-nq_{k,n}e_{k,n}/\ell_n|>\epsilon_n n/\ell_n\}.
}
We establish that $|\Y_k|$ concentrates  in the following lemma.
\begin{restatable}{lemma}{Yconc}
\label{lemma:y_kconc}
For $\epsilon_n > 0$ and n large enough,
\begin{align}
    \Pr(\mathcal{E}_{k,n}) \leq 2e^{-n\epsilon_n^2/(2\ell_n)} + 2e^{-\frac{8 \epsilon_n^2 \ell_n}{(1+2\epsilon_n)} n}.
\end{align}
\end{restatable}
We prove this in the Appendix~\ref{app:Y_rConc}.
The lemma indicates that with high probability, 
$|\Y_k|$ is close to $nq_{k,n}e_{k,n}/\ell_n$.
 We set $\epsilon_n = 1/\log{n}$, ensuring that 
 $\epsilon_n \to 0$ and $\text{Pr}(\mathcal{E}_{k,n}) \to 0$ from Lemma~\ref{lemma:y_kconc}. Then
\begin{align}
    H(\Y_k) &\leq H(\Y_k,\mathbf{1}_{\mathcal{E}_{k,n}}) \leq 1 + H(\Y_k|\mathbf{1}_{\mathcal{E}_{k,n}}) \leq 1 + 2n\text{Pr}(\mathcal{E}_{k,n}) + H(\Y_k|\overline{\mathcal{E}}_{k,n}).
    \label{eq:hyk}
\end{align}
Here, we loosely upper bound $H(\Y_k|\mathcal{E}_{k,n})$ with $2n$ since $\Y_k$ is fully described by $X^n$ and  
$n-1$ 
binary variables that indicate whether there is a tear between the $(i-1)$th and $i$th bits.
We now need to upper bound $H(\Y_k|\overline{\mathcal{E}}_{k,n})$. 
We first note that the total number of possible distinct sequences in $\Y_k$ are
\begin{align}
    \sum_{i=\frac{k-1}{L}\log{n}}^{\frac{k}{L}\log{n}}2^i < 2\times2^{\frac{k}{L}\log{n}} = 2n^{k/L}.
\end{align}
Now given $\overline{\mathcal{E}}_{k,n}$, 
\begin{align}
  |\Y_k| \leq  M_k \triangleq \left(\epsilon_n + q_{k,n}e_{k,n}\right)n/\ell_n.
\end{align}
Following the counting argument in \cite{DNAStorageISIT}, we note that the set $\Y_k$ can be viewed as a histogram over $2n^{k/L}$ sequences. 
Moreover, we can view the last element of the histogram as containing ``excess counts'' if $|\Y_k| < M$, so that the sum of the histogram entries is exactly $M$.
This allows us to bound the term $H(\Y_k|\overline{\mathcal{E}}_{k,n})$ as
\begin{align}
    H & (\Y_k|\overline{\mathcal{E}}_{k,n}) 
    \leq \log\binom{2n^{k/L} + M_k -1}{M_k} \nonumber \leq M_k\log\left(\frac{e(2n^{k/L} + M_k - 1)}{M_k}\right) \nonumber \\
    &= M_k(\log(2n^{k/L} + M_k - 1) + \log{e} -\log{M_k}) \nonumber \\
    &\stackrel{(a)}{=} M_k\left[\max\middle(\frac{k}{L}\log{n},\log{M_k}\middle)
    + \log{e} -\log{M_k} +  P\right] \nonumber \\
    &= M_k\left[\left(\frac{k}{L}\log{n} - \log{M_k}\right)^+ \hspace{-2mm}
    + \log{e} + P\right],
\end{align}
where
$$P := \min\left(\log\middle(2 + \frac{M_k-1}{n^{k/L}}\middle),
    \log\middle(1 + \frac{2n^{k/L}-1}{M_k}\middle)\right).$$
    In step (a), we employ the fact that if $a = b + c = d + e$, then $a = \max(b,d) + \min(c,e)$. Specifically, we can select $(a,b,c,d,e)$ as 
    \begin{align*}
        a := \log(2n^{k/L} + M_k - 1) &= \log{\left(n^{k/L}\left(2+\frac{M_k - 1}{n^{k/L}}\right)\right)} \\
        &= \frac{k}{L}\log{n} + \log{\left(2+\frac{M_k - 1}{n^{k/L}}\right)} := b + c
    \end{align*}
    and
    \begin{align*}
        a:= \log(2n^{k/L} + M_k - 1) &= \log{\left(M_k\left(1+\frac{2n^{k/L} - 1}{M_k}\right)\right)} \\
        &= \log{M_k} + \log{\left(1+\frac{n^{k/L} - 1}{M_k}\right)} := d + e
    \end{align*} 
to arrive at the required expansion.
Proceeding from (\ref{eq:hyk}), this implies that
\begin{align}\label{eq:SetUpperBound}
    &\frac{H(\Y_k)}{n}\leq  \frac{1 + 2n\text{Pr}(\mathcal{E}_{k,n}) + H(\Y_k|\overline{\mathcal{E}}_{k,n})}{n} \nonumber\leq \frac{M_k}{n}\left(\frac{k}{L}\log{n} - \log{M_k}\right)^+
    + A(k,n)\nonumber \\
    &\leq \frac{M_k\log{n}}{n}\left(\frac{k}{L} - \frac{\log{M_k}}{\log{n}}\right)^+ + A(k,n) \stackrel{(a)}{=}  \frac{M_k\log{n}}{n}\left(\frac{k}{L} -  \frac{\log{(n(\epsilon_n + q_{k,n}e_{k,n})/\ell_n)}}{\log{n}}\right)^+ \nonumber \\
    &+ A(k,n) \stackrel{(b)}{\leq} \frac{M_k\log{n}}{n}\left(\frac{k}{L} -  \frac{\log{(n\epsilon_n/\ell_n)}}{\log{n}}\right)^+ + A(k,n) \nonumber \\
    &\stackrel{(c)}{\leq} \frac{M_k\log{n}}{n}\left(\frac{k}{L} - 1\right)^+ + A(k,n) + \frac{M_k\log{n}\log{(\ell_n\log{n})}}{n} \nonumber \\
    &\leq \frac{\log{n}}{\ell_n}(\epsilon_n + q_{k,n}e_{k,n})\left(\frac{k}{L} - 1\right)^+ \hspace{-1mm} + A(k,n) + \frac{(\epsilon_n + 1)\log{(\ell_n\log{n})}}{\ell_n}\log{n}
\end{align}
where
$
A(k,n) \triangleq \frac{1}{n} + 2\text{Pr}(\mathcal{E}_{k,n}) + \frac{M_k}{n}\left(\log{e} + P\right)
$. $(a)$ is due to the definition of $M_k$, $(b)$ is because $q_{k,n}e_{k,n} \geq 0$ and $(c)$ follows because $\epsilon_n = 1/\log{n}$.
This allows us to bound $\sum_{k=1}^{J}\frac{H(\Y_k)}{n}$ as follows:
\begin{align}\label{eqn:FanoFirstBound}
     \sum_{k=1}^{J}\frac{H(\Y_k)}{n} 
    &\stackrel{(a)}{\leq} \frac{\log{n}}{\ell_n}\sum_{k=1}^{J}q_{k,n}e_{k,n}\left(\frac{k}{L}-1\right)^+ + A(n) = \frac{\log{n}}{\ell_n}\sum_{k=L+1}^{J}q_{k,n}e_{k,n}\left(\frac{k}{L}-1\right) + A(n)\nonumber \\
    &\stackrel{(b)}{\leq} \frac{\log{n}}{\ell_n}\sum_{k=L+1}^{\infty}\frac{k}{L}q_{k,n}e_{k,n} +  A(n) 
    - \frac{\log{n}}{\ell_n}E[\mathbf{1}_{\tilde B}],
\end{align}
where in $(a)$ we define 
\begin{align*}
A(n) \triangleq \sum_{k = 1}^{J}\frac{(\epsilon_n + 1)\log{(\ell_n\log{n})\log{(n)}}}{\ell_n} &+ \sum_{k=1}^J\epsilon_n\left(\frac{k}{L}-1\right)^+  + \sum_{i=1}^{J}A(k,n).
\end{align*}
 $(b)$ holds if we recall the definition of the event $\tilde B = \{\vec X_1 \in \Y_{\geq 1}\}$ and note that
$\sum_{k=L+1}^\infty q_{k,n}e_{k,n} = E[\1_{\tilde B}]$. The first term in (\ref{eqn:FanoFirstBound}) is
\begin{align}\label{eq:FirstSumChainRule}
    &\frac{\log{n}}{\ell_n}\sum_{k=L+1}^{\infty}\frac{k}{L}q_{k,n}e_{k,n} = \sum_{k=L+1}^{\infty}\tfrac{q_{k,n}}{\ell_n}E\left[\tfrac k L \log{n}\mathbf{1}_{B}\middle|\tfrac{N_1L}{\log n} \in [k-1,k)\right] \nonumber \\
    &\stackrel{(a)}{=}
    \sum_{k=L+1}^{\infty}\tfrac{q_{k,n}}{\ell_n}E\left[N_1\mathbf{1}_{B}\middle|\tfrac{N_1L}{\log n} \in [k-1,k)\right]  + \sum_{k=L+1}^{\infty}\tfrac{q_{k,n}}{\ell_n}E\left[\delta(N_1)\mathbf{1}_{B}\middle|\tfrac{N_1L}{\log n} \in [k-1,k)\right] \nonumber \\
    &= \frac{E[N_1\mathbf{1}_{\tilde B}]}{\ell_n}  + \sum_{k=L+1}^{\infty}\tfrac{q_{k,n}}{\ell_n}E\left[\delta(N_1)\mathbf{1}_{B}\middle|\tfrac{N_1L}{\log n} \in [k-1,k)\right]
\end{align}
where, in $(a)$, we define $\delta(N_1) \triangleq \frac{k}{L}\log{n} - N_1$. 
Note that given $\frac{k-1}{L}\log{n} \leq N_1 < \frac{k}{L}\log{n}$,
\begin{align}\label{eq:deltabound}
    \delta(N_1) \leq (\log{n})/{L}.
\end{align}
The second summation in \eqref{eq:FirstSumChainRule}
can be upper bounded as
\begin{align}\label{eq:BoundDeltaExp}
     \frac{1}{\ell_n}\sum_{k = L + 1}^{\infty}q_kE[\delta(N_1) \mathbf{1}_{B}| \tfrac{N_1L}{\log n} \in [k-1,k)]&\stackrel{(a)}{\leq} \frac{\log n}{\ell_n} \sum_{k = L + 1}^{\infty} \frac{q_k}{L} E[\mathbf{1}_{B}| \tfrac{N_1L}{\log n} \in [k-1,k)] \nonumber \\
        &\leq \frac{\log{n}}{\ell_nL} \sum_{k = L + 1}^{\infty}q_k 
        \stackrel{(b)}\leq \frac{\log{n}}{\ell_nL},
\end{align}
where (a) follows from (\ref{eq:deltabound}) and (b) follows because $q_{k,n}$ is a probability mass function over $k \in \{0,1,2,\dots\}$.
In summary equations \eqref{eqn:FanoFirstBound}-\eqref{eq:BoundDeltaExp} show that
\begin{align}\label{eq:FinalFirstTerm}
    \sum_{k=1}^{J}\frac{H(\Y_k)}{n} \leq \frac{E[N_1\mathbf{1}_{\tilde B}]}{\ell_n} - \frac{\log{n}}{\ell_n}E[\mathbf{1}_{\tilde B}] + \frac{\log{n}}{\ell_nL} + A(n).
\end{align}
Lemma~\ref{lemma:BigPiecesNo} formalizes $H(\Y_{\geq J})/n$ being finite as $n \to \infty$, and Lemma~\ref{lemma:DistractionTerms} handles the term $A(n)$.
\begin{restatable}{lemma}{BigPiecesNo}
\label{lemma:BigPiecesNo}
The entropy of the set $\Y_{\geq J}$ is upper bounded as
$$
    \lim_{n \to \infty}\frac{H(\Y_{\geq J})}{n} \leq  2\left(S\sqrt{L/J} + \delta\right),
$$
for some finite $S$, and every $J$, $L$ and $\delta > 0$.
\end{restatable}
\begin{restatable}{lemma}{DistractionTerms}
\label{lemma:DistractionTerms} 
As $n \to \infty$, $A(n) \to 0$.
\end{restatable}
The proofs of Lemmas~\ref{lemma:BigPiecesNo} and \ref{lemma:DistractionTerms} are in Appendix~\ref{app:DistractionTerms} and \ref{app:BigPiecesNo}.
From \eqref{eq:Fano} and \eqref{eq:FinalFirstTerm}, we obtain
\begin{align}\label{eq:FinalRateSteps}
    R &\leq \lim_{n \to \infty}\left(\frac{E[N_1\mathbf{1}_{\tilde B}]}{\ell_n} - \frac{\log{n}}{\ell_n}E[\mathbf{1}_{\tilde B}] + A(n) + \frac{H(\Y_{\geq J})}{n}\right) \nonumber \\
    &+ \lim_{n \to \infty}\frac{\log{n}}{\ell_nL}\nonumber \\
    &\stackrel{(a)}{\leq} \Co_{d}\{\log{n}\} - \RC_d\{\log{n}\}  + 2\left(S\sqrt{L/J} + \delta\right)
    + \alpha/ L,
\end{align}
where $(a)$ is due to Lemmas~\ref{lemma:BigPiecesNo} and \ref{lemma:DistractionTerms}, followed by Definitions~\ref{defn:Coverage} and \ref{defn:ReorderingLoss}. Note here that the equivalence between the terms in \eqref{eq:FinalRateSteps} and Definitions~\ref{defn:Coverage} and \ref{defn:ReorderingLoss} are implied by Lemmas~\ref{Lemma:ConcBoundCoverage} and \ref{Lemma:ConcBoundReordering} (which are stated in the previous section).
Further, note that \eqref{eq:FinalRateSteps} holds for all integers $J>L$ and any $\delta > 0$. We can thus pick $L = \log{J}$ and
let $\delta \to 0 $ and $J \to \infty$.
This proves the converse part of Theorem~\ref{thm:Main}.


\section{The Noisy Torn Paper Channel}
\label{Section:TPCLPNoisy}


In the previous section, we characterized the capacity of a TPC with arbitrary fragment length distribution and fragment deletion probabilities. 
The capacity expression provides insights on the impact of strand breaks and missing strands in DNA data storage, but does not incorporate the practical limitation that DNA synthesis and sequencing technologies are subject to symbol-level noise, including substitutions, insertions and deletions.
For that reason, in this section we study the impact of bit-wise noise in the capacity of the TPC.

We consider the TPC with BSC noise as shown in Figure~\ref{fig:TPCNoisy}. Note that this a straightforward extension of the setting discussed in Section~\ref{SubSection:TPCLP-ProbSetting}. The only difference is that the channel first passes the input string through a Binary Symmetric Channel (BSC) with crossover probability $p$. 
The rest of the process remains the same. For simplicity we also assume that no fragments are lost. 
As before we seek to characterize the capacity of this channel.

We conjecture that, as in the noiseless case, the capacity of this channel will retain the form of ``\cov{} $-$ \RC''. But the presence of noise adds a layer of complexity, making it more difficult to express the capacity of this channel in general.
We instead prove inner and outer bounds that match for certain choices of fragment length distributions.
The main theorem of this section is as follows:

\begin{theorem}\label{thm:mainnoisy}
    For the TPC with BSC$(p)$ noise, all rates $R$ less than $R_{\rm in}$ are achievable, where
    \begin{align}\label{eq:NoisyAchi}
        R_{\rm in} = (1-H(p))\cov{\left\{\frac{\log{n}}{1-H(p)}\right\}} -\recost{\left\{\frac{\log{n}}{1-H(p)}\right\}}.
    \end{align}
    Moreover no rates $R$ greater than $R_{\rm out}$ are achievable, where
    \begin{align}
        R_{\rm out} = (1-H(p))\cov{\left\{\log{n}\right\}} -\recost{\left\{\frac{2\log{n}}{1-H(2p)}\right\}}.
    \end{align}
\end{theorem}

\begin{figure}[htb!]
\includegraphics[width=\linewidth]{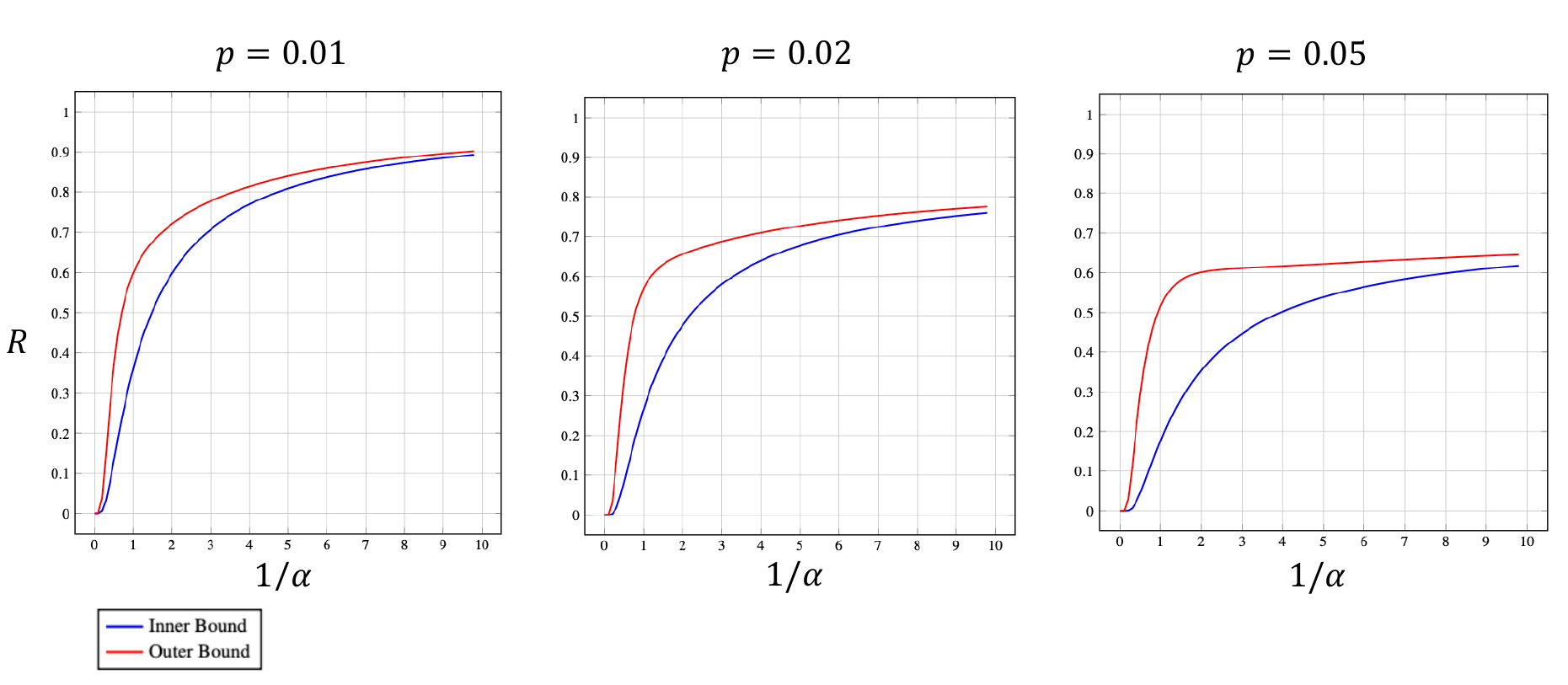}
\caption{\label{fig:TPCcomparison} 
Comparison between inner and outer bounds to the capacity of Noisy \TPC~for $N_1 \sim \text{Geometric}(1/\ell_n)$ and noise parameter (a)$\text{ } p = 0.01$, (b)$\text{ } p = 0.02$ and (c)$ \text{ } p = 0.05$. 
We see that the inner and outer bounds are close to each other as $1/\alpha$ increases and in fact matches when $1/\alpha$ goes to $\infty$.
}
\end{figure}

Notice that if  we consider a fragment length distribution where  $N_i \geq \frac{2\log{n}}{1-H(2p)}$ with probability $1$,
no fragments are discarded in the computation of $\cov{}$ and $\RC{}$ for both the inner bound and the outer bound and the bounds match. 
Specifically the capacity is equal to 
$(1-H(p))\cov{\left\{\frac{\log{n}}{1-H(p)}\right\}} -\recost{\left\{\frac{\log{n}}{1-H(p)}\right\}}$, which matches the capacity conjecture in (\ref{eq:conj}). 
Moreover we notice that even when $N_1$ is allowed to take values less than $\frac{2\log{n}}{1-H(2p)}$, the gap between the lower and upper bounds is not too large as seen in Figure~\ref{fig:TPCcomparison}. Specifically recalling that $1/\alpha = \lim_{n \to \infty} \ell_n/\log{n}$, the x-axis of the figure is the normalized expected length of a fragment. As this value is increased, the gap between the lower and upper bound reduces, as seen in Figure~\ref{fig:TPCcomparison}.

Moreover when $N_1 \geq \frac{2\log{n}}{1-H(2p)}$ with probability $1$, we have $\Co = 1$ and $\RC = \alpha$, where we recall that \(\alpha := \lim_{n \to \infty}\log{n}/\ell_n\). This implies the following corollary which states the capacity for a class of noisy TPCs.

\begin{cor}\label{cor:mainnoisy}
If $N_1 \geq \frac{2\log{n}}{1-H(2p)}$ with probability $1$, the capacity $C_\textrm{noisy-TPC}$ of the TPC is 
\begin{align} \label{eq:corollary_capacity}
    %
    C_\textrm{noisy-TPC} = 1 - H(p) - \alpha.
\end{align}
\end{cor}

Interestingly, this result for fragment lengths greater than the threshold \(\ell_n \geq \frac{2\log{n}}{1-H(2p)}\) matches the capacity of the noisy shuffling channel. 
This implies that, beyond this threshold, the relative differences in fragment lengths become irrelevant. 
For example, two channels—one with an arbitrary distribution where \(E[N_i] = \ell_n\), and the noisy shuffling channel with deterministic fragment lengths \(\ell_n\)—have identical capacities. 
This is surprising since the capacity is independent of the distribution differences between the two channels, when \(N_1 \geq \frac{2\log{n}}{1-H(2p)}\).




\subsection{Achievable rates via typical covering}
\label{SubSection:TPCLPNoisy-AchievableRates}

We again use a random coding argument to prove the achievability of Theorem~\ref{thm:mainnoisy}. But we cannot use a decoder that simply looks for codewords that contain all sequences in $\Y$ as substrings of the codeword, as we did before. This is because these fragments are in general corrupted by BSC noise. Therefore we introduce the concept of \emph{typical covering} instead.


We generate a codebook $\mathcal{C}$ with $2^{nR}$ codewords, by independently picking each letter as $\text{Bern}(1/2)$. 
Let the resulting random codebook be $\mathcal{C} = \{\mathbf{x}_1,\mathbf{x}_2,\dots,\mathbf{x}_{2^{nR}}\}$. 
Assume that $W=1$ is the message that is transmitted. 
The output $\Y$ is available at the decoder. 

We consider a suboptimal decoder that throws away all pieces of size less than $\gamma \log{n}$, where $\gamma$ is a parameter that we will optimize later. We call the new set $\Y_{\geq\gamma}$.
Now we define the notion of \emph{typical covering} on the set $\Y_{\geq\gamma}$ and string $x^n$.
\begin{defn}(Typical covering)
    The set $\Y_{\geq \gamma}$ is said to typically  cover a codeword $x^n$ if the set of reads in $\Y_{\geq \gamma}$ can be aligned to $x^n$, such that the aligned segments are jointly typical. That is for \(\vec Y\), a given ordering of \(\Y_{\geq \gamma}\)
\begin{align}
    \left( x^n, \vec Y \right) \in T_{\epsilon}^n(X, Y),
\end{align}
where \(T_{\epsilon^{\prime}}^n(X, Y)\) denotes the set of \(\epsilon^{\prime}\)-jointly typical sequences \cite{elgamalbook} \((X^n, Y^n)\).
\end{defn}
 The notion of typical covering is essentially a general version of the notion of alignment considered in Section~\ref{SubSection:TPCLP-AchievableRates} for the standard \TPC.
We thus bound the probability of error averaged over all codebook choices as
\begin{align}
    &\text{Pr}(\mathcal{E}) = \text{Pr}(\mathcal{E}|W = 1) \nonumber\\
     &= \text{Pr}\left(\exists \text{ } j\neq1: \Y_{\geq\gamma} \text{ typically covers } x_j\middle|W = 1\right)
\end{align}
We now state two lemmas that are straightforward generalizations of the lemmas in Section~\ref{SubSection:TPCLP-AchievableRates}.

\begin{lemma}\label{Lemma:ConcBoundCoverageNoisy}
For any $\epsilon > 0$, as $n \to \infty$,
\begin{align}\label{eqn:ConcBoundCoverageNoisy}
    \Pr\left(\left|\frac{1}{n}\sum_{i=1}^{K}N_i\mathbf{1}_{\{N_i \geq \gamma\log{n}\}} - \frac{E[N_1\mathbf{1}_{\{N_1 \geq \gamma\log{n}\}}]}{\ell_n}\right| > \frac{E[N_1\mathbf{1}_{\{N_1 \geq \gamma\log{n}\}}]}{\ell_n}\epsilon\right) \to 0.
\end{align}
\end{lemma}
\begin{lemma}\label{Lemma:ConcBoundReorderingNoisy}
For any $\epsilon > 0$, as $n \to \infty$,
\begin{align}\label{eqn:ConcBoundReorderingNoisy}
    \Pr\left(\middle|\sum_{i=1}^{K}\mathbf{1}_{\{N_i \geq \gamma\log{n}\}} - \frac{nE[\mathbf{1}_{\{N_1 \geq \gamma\log{n}\}}]}{\ell_n}\middle|\geq \frac{nE[\mathbf{1}_{\{N_1 \geq \gamma\log{n}\}}]}{\ell_n}\epsilon\right)
    \to 0.
\end{align}
\end{lemma}
Now we let $B_1 = (1+\epsilon)\frac{nE[\mathbf{1}_{\{N_1 \geq \gamma\log{n}\}}]}{\ell_n}$ and $B_2 = (1-\epsilon)\frac{E[N_1\mathbf{1}_{\{N_1 \geq \gamma\log{n}\}}]}{\ell_n}$ and define the event
\begin{align}
    \mathcal{B} = \left\{\sum_{i=1}^{K}\mathbf{1}_{\{\vec{X}_i \in \Y_{\geq\gamma}\}} > B_1\right\}\cup\left\{\frac{1}{n}\sum_{i=1}^{K}N_i\mathbf{1}_{\{\vec{X}_i \in \Y_{\geq\gamma}\}} < B_2\right\}.
\end{align}
Lemmas~\ref{Lemma:ConcBoundCoverageNoisy} and \ref{Lemma:ConcBoundReorderingNoisy} imply that $\text{Pr}(\mathcal{B}) \to 0$ as $n \to \infty$. Therefore,
\begin{align}\label{eq:ProbUpperBoundNoisy}
    \text{Pr}(\mathcal{E}) &= \text{Pr}(\exists \text{ } j\neq1: \Y_{\geq\gamma} \text{ typically covers } x_j|W = 1) \nonumber \\
    &\leq\text{Pr}(\exists \text{ } j\neq1: \Y_{\geq\gamma} \text{ typically covers } x_j|W = 1,\overline{\mathcal{B}}) \nonumber \\
    &+ \text{Pr}(\mathcal{B}) \nonumber \\
    &\stackrel{(a)}\leq |\mathcal{C}|\frac{n^{B_1}}{2^{nB_2I(A;B)(1-\epsilon^{\prime})}} + \text{Pr}(\mathcal{B})\nonumber \\
    &\leq 2^{nR}2^{B_1\log{n}}2^{-nB_2(1-H(p))(1-\epsilon^{\prime})} + o(1),
\end{align}
where $A$ is a Bern$\left(\frac{1}{2}\right)$ random variable and $B = A \oplus\text{Bern}(p)$. Inequality $(a)$ follows from the union bound and the fact that given $\overline{\mathcal{B}}$, there are at most $n^{B_1}$ ways to align $\Y_{\geq\gamma}$ to a codeword $x_j$. 
To see this note that, given $|\Y_{\geq\gamma}| < B_1$, there are at most $n$ places the fragments can start from to align each piece and at most $B_1$ such pieces. 
Finally, the probability that there is an erroneous typical covering of length $nB_2$ corresponds to the probability that two independent Bern$\left(\frac{1}{2}\right)$ sequences seem jointly typical according to $(A,B)$, which by standard typicality arguments, 
is at most $2^{-nB_2I(A;B)(1-\epsilon^{\prime})}$ \cite{elgamalbook}. Note here that $I(A;B) = H(A) - H(A|B) = 1 - H(p).$ From this
 we conclude that $\Pr(\mathcal{E}) \to 0$ if
\begin{align}\label{eq:UpperBoud}
    R <  \lim_{n \to \infty} \left(\frac{(1-\epsilon^{\prime})(1-\epsilon)(1-H(p))}{\ell_n}E\left[N_1\mathbf{1}_{N_1\geq\gamma\log{n}}\right] - \frac{(1+\epsilon)\log{n}}{\ell_n}E\left[\mathbf{1}_{N_1\geq\gamma\log{n}}\right]\right). 
\end{align}
Now if we let $\epsilon \to 0$ and $\epsilon^{\prime} \to 0$, we can rewrite \eqref{eq:UpperBoud} as
\begin{align}
    R & < \lim_{n \to \infty}\left( \frac{\log{n}}{\ell_n}E\left[\left(\frac{N_1(1-H(p))}{\log{n}}-1\right)\mathbf{1}_{N_1\geq\gamma\log{n}}\right] \right) \nonumber \\
    &= \lim_{n \to \infty}\left( \frac{\log{n}}{\ell_n}\sum_{x \geq \gamma\log{n}}\left(\frac{x(1-H(p))}{\log{n}}-1\right)p_{N_1}(x) \right).
\end{align}
Note that since $p_{N_1} \geq 0$, 
the best achievability is obtained by setting $\gamma$  to be the least value such that $\left(\frac{x(1-H(p))}{\log{n}}-1\right)$ is greater than $0$, which maximizes the above expression. This value is $\gamma^{*} = \frac1{1-H(p)}$.
Substituting this value of $\gamma$, we conclude that all rates $R \leq R_{\rm in}$ 
are achievable.
\subsection{Outer-bound}
\label{SubSection:TPCLPNoisy-Converse}

In order to prove the converse for the noisy case, we again partition the set $\Y$ into a set of parallel channels, each with fragments of roughly the same length.
To each of these channels, we apply a generalization of the result in \cite{noisyshuffling}, which considered a noisy shuffling channel with pieces of a fixed size. 
However, the result from \cite{noisyshuffling} cannot be applied to the channels with very short fragments.
We therefore need to apply a different, possibly looser bound for the channels with short fragments.
More precisely let $\vec Y_i$ be the output of the string $\vec X_i$ after it is passed through the BSC$(p)$ channel. Then we define $\Y_k$ as
\begin{align}\label{eq:y_k_new_def}
    \Y_k \triangleq \left\{\vec{Y}_i \in \Y: \frac{k-1}{L}\log{n} \leq N_{i} < \frac{k}{L}\log{n}\right\}.
\end{align}
Note that the above definition is slightly different from the definition used in equation \eqref{eq:y_kdef} from Section~\ref{SubSection:TPCLP-Converse}. Since fragments are corrupted by noise, the output is not just the unordered set of input strings.
Then, from a standard argument with Fano's inequality, we have
\begin{align}\label{eq:FanoNoisy}
    R &\leq \lim_{n \to \infty} \frac{I(X^n;\Y)}{n}.
\end{align}
 Let $Z^n$ be a binary string representing whether a bit has been flipped at a given position in $X^n$. We also define $T_2^n$ to be the sequence of tearing locations on string $X^n$. More precisely, let $T_2,T_3,\dots,T_n$ be binary indicator random variables of whether there is a cut between $X_{i-1}$ and $X_i$.
The random vector 
$\vec{\Z}_k$ is the ordered set of substrings of size in $\left[\frac{k-1}{L}\log{n},\frac{k}{L}\log{n}\right)$ extracted from $Z^n$ based on tearing locations $T_2^n$. That is, if we take two successive $1$s in $T_2^n$, say $T_i = 1$ and $T_j = 1$, such that $j-(i+1) \in \left[\frac{k-1}{L}\log{n},\frac{k}{L}\log{n}\right)$, then $Z_{i+1}^{j} \in\vec Z_k$. $\vec{\X}_k$ is defined in a similar way.

 We further define $\Y_{\leq A} := \{\vec{Y}_i:N_i \leq (A/L)\log{n}\}$, for some constant $A$. 
 The definitions of $\Y_{\in [A:B]}$ and $\Y_{\geq A}$ are similar. Then for an integer $J$, we have that
\begin{align}\label{eq:FanoExpansion}
    &I(X^n;\Y) = H(\Y) - H(\Y|X^n) \leq H(\Y_{\leq L}) + H(\Y_{\in  (L:J)}) + H(\Y_{\geq J}) - H(\Y|X^n,T_2^n) \nonumber \\
    &= H(\Y_{\in  (L:J)}) - H(\Y,Z^n|X^n,T_2^n) + H(Z^n|X^n,\Y,T_2^n) + H(\Y_{\leq  L}) +  H(\Y_{\geq J})\nonumber \\
    &=  H(\Y_{\in  (L:J)}) + H(Z^n|X^n,\Y,T_2^n) - nH(p) + H(\Y_{\leq L}) + H(\Y_{\geq J}).
\end{align}
Now to bound the term $H(Z^n|X^n,\Y,T_2^n)$, we  note that, given the tearing locations $T_2^n$, $Z^n$ is a function of $\vec{Z}_1,\vec{Z_2},\dots$, since the substrings in each $\vec{\Z}_k$ (which is an ordered set) can be uniquely located in $Z^n$ based on the tearing points $T_2^n$. Hence,
\begin{align}\label{eq:tearingentropy}
    &H(Z^n|X^n,\Y,T_2^n) \leq H(\vec \Z_1, \vec \Z_2, \dots, \vec \Z_J, \vec \Z_{\geq J} |X^n,\Y,T_2^n) \\ \nonumber 
    &\leq \sum_{k=1}^{J}H(\vec{\Z}_k|X^n,\Y) + H(\vec{\Z}_{\geq J}) \leq \sum_{k=1}^LH(\vec{\Z}_k) + \sum_{k=L+1}^{J} H(\vec{\Z}_k|X^n,\Y) + H(\vec{\Z}_{\geq J}).
\end{align}
 Therefore from \eqref{eq:FanoExpansion} and \eqref{eq:tearingentropy}, and using this definition one can conclude that
\begin{align}
    I(X^n;\Y) &\leq \sum_{k = L+1}^{J}\left(H(\Y_{k}) + H(\vec{\Z}_{k}|\vec{\X}_{k},\Y_{k})\right) + \sum_{k=1}^{L}H(\vec{\Z}_{k}) \nonumber \\
    &- nH(p) + H(\Y_{\leq L}) + H(\Y_{\geq J}) + H(\vec{\Z}_{\geq J}).
\end{align}
Now we follow steps similar to Section~\ref{SubSection:TPCLP-Converse}. First we define the probability
\begin{align}
    q_{k,n} &= \text{Pr}\left(\frac{k-1}{L}\log{n} \leq N_1 < \frac{k}{L}\log{n}\right).
\end{align}
Additionally, we define the event 
\aln{
\mathcal{E}_{k,n} =
\{||\Y_k|-nq_{k,n}/\ell_n|> n\epsilon_n/\ell_n\}.
}
We establish that $|\Y_k|$ concentrates in the following lemma, which is similar to Lemma~\ref{lemma:y_kconc}, but for the \(\Y_k\) defined in \eqref{eq:y_k_new_def}.
\begin{lemma}\label{lemma:y_kconcnoisy}
For $\epsilon_n > 0$ and n large enough,
\begin{align}
    \Pr(\mathcal{E}_{k,n}) \leq 2e^{-n\epsilon_n^2/(2\ell_n)} + 2e^{-\frac{8 \epsilon_n^2 \ell_n}{(1+2\epsilon_n)} n}.
\end{align}
\end{lemma}
The proof of the lemma is the similar to Lemma~\ref{lemma:y_kconc}. As before we can set $\epsilon_{n} = 1/\log{n}$ to guarantee that $\epsilon_{n} \to 0$ and $\Pr(\mathcal{E}_{k,n}) \to 0$, as $n \to\infty$ for a fixed $k$. 
Now, given $\bar{\mathcal{E}}_{k,n}$,
\begin{align}\label{eq:upperlowercard}
     \left(-\epsilon_{n} +q_{k,n}\right)\frac{n}{\ell_n}\leq |\vec \Z_{k}| = |\vec \X_{k}| = |\Y_{k}| \leq M_{k},
\end{align}
where
\begin{align}
    M_{k} := \left(\epsilon_{n} +q_{k,n}\right)\frac{n}{\ell_n}.
\end{align}
    As in Section~\ref{SubSection:TPCLP-Converse}, we expand the terms $H(\Y_k)$ and $H(\vec \Z_k|\vec \X_k,\Y_k)$ as  
\begin{align}\label{eq:H(Y_r)UpperBd}
   &H(\Y_k) + H(\vec \Z_k|\vec \X_k,\Y_k) \leq H(\Y_k,\mathbf{1}_{\mathcal{E}_{k,n}}) + H(\vec \Z_k,\mathbf{1}_{\mathcal{E}_{k,n}} |\vec \X_k,\Y_k) \nonumber \\
   &\leq 2(H(\Pr(\mathcal{E}_{k,n})) + 2n\Pr(\mathcal{E}_{k,n})) + H(\Y_k|\bar{\mathcal{E}}_{k,n}) + H(\vec \Z_k|\vec \X_k,\Y_k,\bar{\mathcal{E}}_{k,n}).
\end{align}
Let us first consider the term $H(\Y_{\leq L})$ which appears in \eqref{eq:FanoExpansion}.  The term $H(\Y_{\leq L})/n$ can be expanded as follows
\begin{align}\label{eq:H(Y_s)UpperBd}
    &\frac{H(\Y_{\leq L})}{n} = \sum_{k=1}^{L}\frac{H(\Y_{k})}{n} \stackrel{(a)}{\leq} \sum_{k=1}^{L}\left(\frac{H(\Y_{k}|\bar{\mathcal{E}}_{k,n})}{n} + 2\Pr(\mathcal{E}_{k,n}) + \frac{1}{n}\right),
\end{align}
where $(a)$ follows similarly to \eqref{eq:H(Y_r)UpperBd}. 
We can thus bound $I(X^n;\Y)/n$ using \eqref{eq:FanoExpansion}, \eqref{eq:H(Y_r)UpperBd},  \eqref{eq:H(Y_s)UpperBd} as
\begin{align}\label{eq:termsplitup}
    \frac{I(X^n;\Y)}{n} &\leq \frac{1}{n}\sum_{k=L+1}^{J}\left(H(\Y_{k}|\bar{\mathcal{E}}_{k,n}) + H(\vec{\Z}_{k}|\vec{\X}_{k},\Y_{k},\bar{\mathcal{E}}_{k,n})\right) + \frac{1}{n}\sum_{k=1}^{L}H(\Y_{k}|\bar{\mathcal{E}}_{k,n}) \nonumber \\
    &+ 2\sum_{k=1}^{J}\left(2\Pr(\mathcal{E}_{k,n}) + \frac{1}{n}\right) + \frac{H(\Y_{\geq J}) + H(\vec{\Z}_{\geq J})}{n} + \frac{1}{n}\sum_{k=1 }^{L}H(\vec{\Z}_{k}|\bar{\mathcal{E}}_{k,n}) - H(p) .
\end{align}
Notice the first term in \eqref{eq:termsplitup}. They contain entropy terms on pieces of ``roughly'' the same size.
$H(\Y_{k}|\bar{\mathcal{E}}_{k,n})$ and $H(\vec{\Z}_{k}|\vec{\X}_{k},\Y_{k},\bar{\mathcal{E}}_{k,n})$  contend with each other in a way similar to \cite{noisyshuffling}. 
Carefully using this tension lets us apply a non-trivial upper bound on the sum of these entropies in a certain regime. 
This is summarized formally in the following lemma, which is the main work-horse to prove the outer bound stated in Theorem~\ref{thm:mainnoisy}.
\begin{restatable}{lemma}{NoisyUpperBound}
\label{lem:NoisyUpperBound}
If $\frac {k-1}{L} > \frac{2}{1 - H(2p)}$, we have
\begin{align}\label{eq:Yconditioning}
    &H(\Y_{k}|\bar{\mathcal{E}}_{k,n}) + H(\vec{\Z}_{k}|\vec{\X}_{k},\Y_{k},\bar{\mathcal{E}}_{k,n}) \nonumber \\
    &\leq \frac{\log^2{(n+1)}}{L} + M_k\frac{k}{L}\log{n}- \left(M_k-\frac{2n\epsilon_n}{\ell_n}\right)\log\left(M_k - \frac{2n\epsilon_n}{\ell_n}\right)  + o(M_k\log{M_k}),
\end{align}
else
\begin{align}\label{eq:Zconditioning}
    H(\Y_{k}|\bar{\mathcal{E}}_{k,n}) + H(\vec{\Z}_{k}|\vec{\X}_{k},\Y_{k},\bar{\mathcal{E}}_{k,n})\leq \frac{\log^2{(n+1)}}{L} + M_k\frac{k}{L}\log{n} + o(M_k\log{M_k}).
\end{align}
\end{restatable}
The lemma above is the main additional technical step in this proof compared to the noise-free case.
Lemma~\ref{lem:NoisyUpperBound} is proved in Appendix~\ref{app:NoisyUpperBound}. Using this lemma we can evaluate a closed-form expression as an upper bound to the first term in \eqref{eq:termsplitup}. This is summarized in the following lemma:
\begin{restatable}{lemma}{NoisyTerms}\label{lem:NoisyTermsConsolidation}
For $n$ large enough, it holds that
\begin{align}
    &\sum_{k= L+1}^{J}\left(H(\Y_k|\bar{\mathcal{E}}_{k,n}) + H(\vec{\Z}_k|\vec{\X}_k,\Y_k,\bar{\mathcal{E}}_k)\right) \nonumber \\
    &= n\left(\frac{E\left[N_1\mathbf{1}_{\{N_1\geq\log{n}\}}\right]}{\ell_n} - \frac{\log{n}}{\ell_n}E\left[\mathbf{1}_{\left\{N_1\geq\frac{2\log{n}}{1-H(2p)}\right\}}\right]\right) + o(n).
\end{align}
\end{restatable}
Lemma~\ref{lem:NoisyTermsConsolidation} is proved in Appendix~\ref{app:NoisyTermsConsolidation}. We now a state some lemmas to upper bound the other terms in \eqref{eq:termsplitup}.
Lemma~\ref{lemma:BigPiecesNoNoisy} below formalizes $(H(\Y_{\geq J}) + H(\vec{Z}_{\geq J}))/n$ being finite as $n \to \infty$.
\begin{restatable}{lemma}{BigPiecesNoNoisy}
\label{lemma:BigPiecesNoNoisy}
The entropy of the set $\Y_{\geq J}$ is upper bounded as
$$
    \lim_{n \to \infty}\frac{H(\Y_{\geq J})+H(\vec{\Z}_{\geq J})}{n} \leq  4\left(S\sqrt{L/J} + \delta\right),
$$
for some finite $S$, and every $J$, $L$ and $\delta > 0$.
\end{restatable}
This lemma is a straightforward extension of Lemma~\ref{lemma:BigPiecesNo}. Moreover we can prove the following lemma to show that the second and third summations in \eqref{eq:termsplitup} are small as $n \to \infty$ .
\begin{restatable}{lemma}{NoisyDistraction}
\label{lem:noisydistraction}
Define
\begin{align}
B(n):= \frac{1}{n}\sum_{k=1}^{L}H(\Y_{k}|\bar{\mathcal{E}}_{k,n}) +  2\sum_{k=1}^{J}\left(2\Pr(\mathcal{E}_{k,n}) + \frac{1}{n}\right).
\end{align}
Then $B(n) \to 0$ as $n \to \infty$.
$$\lim_{n\to\infty}B(n) = 0.$$
\end{restatable}
This lemma is proved in Appendix~\ref{app:noisydistraction}. We can thus conclude by applying Lemma~\ref{lem:NoisyTermsConsolidation}, Lemma~\ref{lem:noisydistraction} and Lemma~\ref{lemma:BigPiecesNoNoisy} to \eqref{eq:termsplitup} that
\begin{align}
    &R \leq \lim_{n \to \infty} \frac{I(X^n;\Y)}{n} \stackrel{(a)}{\leq} \lim_{n \to \infty}\frac{E\left[N_1\mathbf{1}_{\{N_1\geq\log{n}\}}\right]}{\ell_n} - \lim_{n \to \infty}\frac{\log{n}}{\ell_n}E\left[\mathbf{1}_{\left\{N_1\geq\frac{2}{1-H(2p)}\log{n}\right\}}\right] \nonumber \\ 
    &+ \lim_{n\to\infty} B(n) + 4\left(S\sqrt{L/J} + \delta\right)- H(p) + \lim_{n \to \infty}\frac1n\sum_{k=1}^{L}H(\vec{\Z}_{k}|\overline{\mathcal{E}}_{k,n}) + \lim_{n\to\infty} o(1)\nonumber \\
    &\stackrel{(b)}{\leq} \lim_{n \to \infty}\frac{E\left[N_1\mathbf{1}_{\{N_1\geq\log{n}\}}\right]}{\ell_n} - \lim_{n \to \infty}\frac{\log{n}}{\ell_n}E\left[\mathbf{1}_{\left\{N_1\geq\frac{2}{1-H(2p)}\log{n}\right\}}\right] - \lim_{n \to \infty}\frac{H(p)}{\ell_n} E[N_1] \nonumber \\
    &+ \lim_{n \to \infty}\frac{H(p)}{\ell_n}\sum_{k=1}^L(\epsilon_n + q_{k,n})\frac{k}{L}\log{n} + 4\left(S\sqrt{L/J} + \delta\right)  \nonumber \\
    &\stackrel{(c)}{\leq} \lim_{n \to \infty}\frac{E\left[N_1\mathbf{1}_{\{N_1\geq\log{n}\}}\right]}{\ell_n} - \lim_{n \to \infty}\frac{\log{n}} {\ell_n}E\left[\mathbf{1}_{\left\{N_1\geq\frac{2}{1-H(2p)}\log{n}\right\}}\right] \nonumber \\
    &+ \frac{H(p)}{\ell_n}E[N_1\mathbf{1}_{\{N_1\geq \log{n}\}}] + \lim_{n \to \infty}\left(\frac{\log{n}}{\ell_nL} +\frac{LH(p)\epsilon_n}{\ell_n}\right) + 4\left(S\sqrt{L/J} + \delta\right) \nonumber \\
    &=  \lim_{n \to \infty} \left((1-H(p))\frac{E\left[N_1\mathbf{1}_{\{N_1\geq\log{n}\}}\right]}{\ell_n} - \frac{\log{n}}{\ell_n}E\left[\mathbf{1}_{\left\{N_1\geq\frac{2\log{n}}{1-H(2p)}\right\}}\right]\right) + 4\left(S\sqrt{L/J} + \delta\right) + \frac{\alpha}{L}, \nonumber \\
    &= (1-H(p))\Co\{\log{n}\} - \RC\left\{\frac{2\log{n}}{1-H(2p)}\right\} + 4\left(S\sqrt{L/J} + \delta\right) + \frac{\alpha}{L},
\end{align}
%
where $(a)$ is due to Lemma~\ref{lem:NoisyTermsConsolidation} and Lemma~\ref{lemma:BigPiecesNo}, $(b)$ is due to Lemma~\ref{lem:noisydistraction} and the fact that the number of bits in an ordered set $\vec{\Z}_k$ is upper bounded by $$M_k\times\frac{k}{L}\log{n}:= (\epsilon_n + q_{k,n})\frac {k}{L} \times \frac{n}{\ell_n} \log{n},$$ and $(c)$ is due to to arguments considered in equations \eqref{eq:FirstSumChainRule}, \eqref{eq:deltabound} and \eqref{eq:BoundDeltaExp}.
As before we set $L = \log{J}$ and let $J \to \infty$ and $\delta \to 0$.
This proves the converse of Theorem~\ref{thm:mainnoisy}.


\section{Conclusion}
\label{Section:Conclusion}

Motivated by the application of DNA and macromolecular data storage, we have studied the  impact of fragment length distribution, deletion probabilities, and bit-level noise in the capacity of the torn-paper channel.
We have shown that the capacity of a generalized TPC can be expressed as ``\Co$-$\RC'' in the absence of bit-level noise. Moreover we showed inner and outer bounds, again of the form ``\Co$-$\RC'', when the observed sequences are subject to binary symmetric noise.


A natural question for follow-up work is whether the inner and outer bounds on the capacity of the TPC with bit-level noise can be tightened.
We conjecture that the inner bound in Section~\ref{SubSection:TPCLPNoisy-AchievableRates} is tight, and the slack is in the converse argument.
Related results on the Noisy Shuffling Channel \cite{WeinMerhavNoisy} imply that the first bound in Lemma~\ref{lem:NoisyUpperBound} (equation~\ref{eq:Yconditioning}) likely holds given a less stringent condition, which would lead to a tighter outer bound. 
For instance if it were true that (\ref{eq:Yconditioning}) in Lemma~\ref{lem:NoisyUpperBound} holds for any $(p,r)$ such that
$1 - H(p) - 1/r \geq 0$,
instead of the original condition, it can be verified that the same analysis used in Section~\ref{SubSection:TPCLPNoisy-Converse} yields an outer bound that matches the inner bound.

An other natural question, is what the results generalise to when you consider a two dimensional codeword that is torn. This is particularly relevant in forensics where typical messages are encoded in two dimensions (for example in fingerprinting). Precisely this model  is as follows: Assume messages are to be written on a two dimensional codeword $X^{n \times m}$. This codeword is then passed through a torn paper channel that tears this codeword into arbitrary shapes. These pieces are then available to the decoder. What is the capacity of this channel? This channel can be viewed as a generalisation of the TPC, since if we set $m = 1$ we get back the original TPC. But for $m$ greater than $1$, unlike the TPC, spatial information can play a role in discerning the order. Since the shapes are more arbitrary, it may be easier to order pieces of the codeword based on matching shapes, not unlike a ``jigsaw'' puzzle. How we make use of this spatial information or if it is beneficial to do so from a capacity standpoint is subject to future work.


\newpage

\bibliographystyle{unsrtnat}
\bibliography{main}
\newpage
\appendices
\section{Proof of Corollary~\ref{thm:Main2}}\label{app:cor2}
\thmMain*
Corollary~\ref{thm:Main2} provides us with an expression to numerically compute the \TPC{} capacity for certain classes of fragment length distributions ($N_i$). Intuitively it holds for distributions of $N_i$ which have a continuous analog (for example the geometric or uniform distributions). We proceed as follows:
\begin{align}
    &C = \Co_{d}\{\log{n}\} - \RC_{d}\{\log{n}\} \nonumber \\
    &\stackrel{(a)}{=}\lim_{n \to \infty} \left(\frac{E[N_1\mathbf{1}_{\tilde B}]}{\ell_n} - \frac{\log{n}E[\mathbf{1}_{\tilde B}]}{\ell_n}\right) \nonumber \\
    &= \lim_{n \to \infty} (\log{n}/\ell_n) \lim_{n \to \infty} E\left[\left(\frac{N_1}{\log{n}}-1\right)\mathbf{1}_{\tilde B}\right] \nonumber \\
    &= \alpha \lim_{n \to \infty} E\left[\left(\frac{N_1}{\log{n}} - 1\right)E\left[\mathbf{1}_{\tilde B}|N_1\right]\right] \nonumber \\
    &= \alpha\lim_{n \to \infty}\sum_{x = \log{n}}^{\infty}\left(\frac{x}{\log{n}} - 1\right)(1 - \tilde{d}(x))\text{Pr}(N_1 = x).
\end{align}
Note in $(a)$ we use an equivalent definition of $\Co$ and $\RC$ that we employed in \eqref{eq:AchFinalSteps}.
Now we substitute $x = \beta\log{n}$ in the above equation, where $\beta \in \{1, 1 + \frac{1}{\log{n}}, 1 + \frac{2}{\log{n}},\dots\} \triangleq \mathcal{Q}$. 
Then we have
\begin{align}\label{eqn:sumtoint}
   &C = \alpha\lim_{n \to \infty}\sum_{\beta \in  \mathcal{Q}}\left(\beta - 1\right)(1 - \tilde{d}(\beta\log{n}))\text{Pr}(N_1 = \beta\log{n}) \nonumber \\
   &= \alpha\lim_{n \to \infty}\sum_{\beta \in  \mathcal{Q}}\left(\beta - 1\right)(1 - \tilde{d}(\beta\log{n}))\frac{\text{Pr}(N_1 = \beta\log{n})}{\Delta_n}\Delta_n \nonumber \\
   &\stackrel{(a)}{=} \alpha\int_{1}^{\infty}(\beta - 1)(1 - \hat{d}(\beta))h(\beta)d\beta,
\end{align}
where $\lim_{n \to \infty}\text{Pr}(N_1 = \beta\log{n})\log{n} \triangleq h(\beta)$ and $\Delta_n = \beta +  1/\log{n} - \beta$.
$(a)$ follows from the definition of Riemann integration. This result holds when $h(\beta)$ exists and is finite.
\section{Proof of Lemma~\ref{Lemma:ConcBoundCoverage}}\label{app:ConcBoundCov}
\CoverageLemma*
Define $Z_i = N_i\textbf{1}_{\{\vec{X}_i \in \Y_{\geq 1}\}}$, $Z = \sum_{i = 1}^KV_i$ and $\tilde{V} = \sum_{i = 1}^{n/\ell_n}V_i$. Note that $\mathbf{E}[\tilde{V}] = n\mathbf{E}[V_i]/\ell_n$.\\
If $\tilde{V} > 
V$, then $K < n/\ell_n$ and
\begin{align}
    |V - \tilde{V}| \leq \sum_{i = K + 1}^{n/\ell_n}V_i \leq \sum_{i = K + 1}^{n/\ell_n}N_i \leq \left|\sum_{i = 1}^{n/\ell_n}N_i - n\right|.
\end{align}
Similarly if $Z > \tilde{Z}$, then $K > n/\ell_n$ and
\begin{align}
    |V - \tilde{V}| \leq \sum_{i = n/\ell_n}^{K}V_i \leq \sum_{i = n/\ell_n}^K V_i \leq \left|\sum_{i = 1}^{n/\ell_n}N_i - n\right|.
\end{align}
Therefore
\begin{align}
    &\text{Pr}(|V - \tilde{V}| \geq \delta n ) \leq \text{Pr}\left(\left|\sum_{i = 1}^{n/\ell_n}N_i - n\right| \geq \delta n\right) \nonumber \\
    &= \text{Pr}\left(\sum_{i = 1}^{n/\ell_n}N_i \geq n(1 + \delta/\ell_n)\right) + \text{Pr}\left(\sum_{i = 1}^{n/\ell_n}N_i \leq n(1 - \delta/\ell_n)\right) \nonumber \\
    &= \text{Pr}\left(\frac{\ell_n}{n}\sum_{i=1}^{n/\ell_n}(N_i - \mathbf{E}[N_i]) > \delta\right) \nonumber \\
   &+ \text{Pr}\left(\frac{\ell_n}{n}\sum_{i=1}^{n/\ell_n}(N_i - \mathbf{E}[N_i]) < -\delta\right) \leq 2e^{-2\delta^2n/\ell_n}.
\end{align}
Now we use the Chebyshev's inequality to bound $|\tilde{V} - \mathbf{E}[\tilde{V}]|$.
\begin{align}
    &\text{Pr}(|\tilde{V} - \mathbf{E}[\tilde{V}]| > \delta n) \leq \frac{\text{Var}(\tilde{V_1})}{\delta^2n} \leq \frac{\text{E}(V_1^2)}{\delta^2n} \nonumber \\
    &\leq \frac{\mathbf{E}[N_1^2]}{\delta^2n}.
\end{align}
Now if $|V - \tilde{V}| < \frac{n\mathbf{E}[N_1\mathbf{1}_{\tilde B}]}{2}\epsilon $ and $|\tilde{V} - \mathbf{E}[\tilde{V}]| < \frac{n \mathbf{E}[N_1\mathbf{1}_{\tilde B}]}{2}\epsilon$, then by triangle inequality $|V - \mathbf{E}[\tilde{V}]| < n \mathbf{E}[N_1\mathbf{1}_{\tilde B}] \epsilon$. Therefore (define $\epsilon^{\prime} := \mathbf{E}[N_1\mathbf{1}_{\tilde B}]\epsilon$)
\begin{align}
    &\text{Pr}(|V - \mathbf{E}[\tilde{V}]|> n \epsilon^{\prime}) \nonumber \\
    &\leq \text{Pr}(|V - \tilde{V}| > n \epsilon^{\prime}/2) + \text{Pr}(|\tilde{V} - \mathbf{E}[\tilde{V}]| > n \epsilon^{\prime}/2) \nonumber \\
    &\leq 2e^{-{\epsilon^{\prime}}^2 n/(2\ell_n)} + \frac{\mathbf{E}[N_1^2]}{{{\epsilon^{\prime}}^2}n} \sim  2e^{-\epsilon^2n\log^2{n}/(2\ell_n)} +  \frac{1}{\epsilon^2n}  = o(1) \to 0,
\end{align}
as $n \to \infty$.
\newpage
\section{Proof of Lemma~\ref{Lemma:ConcBoundReordering}}\label{app:RCLem}
\ConcBound*
Define ${\epsilon^{\prime}}:=\epsilon E[\mathbf{1}_{\tilde B}]$. We follow steps similar to Lemma~\ref{lemma:y_kconc}.
Let $V_i = \mathbf{1}_{\tilde{B}}$ for $i = 1,2,\dots$ and $\tilde{V} = \sum_{i=1}^{n/\ell_n}V_i$.
We know that,
\begin{align}
    E[\tilde{V}] = \frac{nE[\mathbf{1}_{B'}]}{\ell_n}.
\end{align}
Now if $|\tilde{V} - E[\tilde{V}]| < n{\epsilon^{\prime}}/(2\ell_n)$ and $\left|\sum_{i=1}^{K}\mathbf{1}_{\{\vec{X}_i \in \Y_{\geq 1}\}} - \tilde{V}\right| \leq |K - n/\ell_n| < n{\epsilon^{\prime}}/(2\ell_n)$, by triangle inequality $\left|\sum_{i=1}^{K}\mathbf{1}_{\{\vec{X}_i \in \Y_{\geq 1}\}} - nE[\mathbf{1}_{B'}]/\ell_n\right| \leq n{\epsilon^{\prime}}/(2\ell_n)$. Therefore,
\begin{align}
    &\text{Pr}\left(\left|\sum_{i=1}^{K}\mathbf{1}_{\{\vec{X}_i \in \Y_{\geq 1}\}} - \frac{nE[\mathbf{1}_{B'}]}{\ell_n}\right| \geq n{\epsilon^{\prime}}/\ell_n\right) \leq \text{Pr}\left(|K - n/\ell_n| > n{\epsilon^{\prime}}/(2\ell_n)\right) \nonumber \\
    &+ \text{Pr}\left(\left|\sum_{i=1}^{K}\mathbf{1}_{\{\vec{X}_i \in \Y_{\geq 1}\}} - \frac{nE[\mathbf{1}_{B'}]}{\ell_n}\right| \geq n{\epsilon^{\prime}}/(2\ell_n)\right) \nonumber \\
    &\stackrel{(a)}{\leq} 2e^{-\frac{ {\epsilon^{\prime}}^2 \ell_n}{(2+{\epsilon^{\prime}})} n} + 2e^{-n{\epsilon^{\prime}}^2/(2\ell_n)} \to 0,
\end{align}
uas $n \to \infty$. $(a)$ follows from \eqref{eq:BoundK} and Hoeffding's inequality.
\newpage
\section{Proof of Lemma~\ref{lemma:y_kconc}}\label{app:Y_rConc}
\Yconc*
We first prove a concentration bound for $K$ and utilize that to prove the lemma.
\begin{align}\label{eq:BoundK}
    &\text{Pr}(|K - n/\ell_n| > \delta n/\ell_n)
     = \text{Pr}(K  > (1+\delta) n/\ell_n) + \text{Pr}(K < (1-\delta) n/\ell_n)\nonumber \\
    & \leq \text{Pr}\left( \sum_{i=1}^{{(1+\delta)n/\ell_n}} N_i  \leq  n \right) + \text{Pr}\left( \sum_{i=1}^{ {(1-\delta)n/\ell_n}} N_i  \geq  n \right) \nonumber \\
    & = \text{Pr}\left( \sum_{i=1}^{ {(1+\delta)n/\ell_n}} (N_i  - E[N_i]) \leq  -\delta n \right) + \text{Pr}\left( \sum_{i=1}^{ {(1-\delta)n/\ell_n}} (N_i  - E[N_i]) \geq \delta n \right) \nonumber \\
    & = \text{Pr}\left( \frac{\ell_n}{(1+\delta)n}\sum_{i=1}^{ (1+\delta)n/\ell_n} (N_i  - E[N_i]) \leq  -\frac{\delta \ell_n}{1+\delta} \right) \nonumber \\
    &+ \text{Pr}\left( \frac{\ell_n}{(1-\delta)n}\sum_{i=1}^{ (1-\delta)n/\ell_n} (N_i  - E[N_i]) \geq  \frac{\delta \ell_n}{1-\delta} \right) \nonumber \\
    &\leq e^{-\frac{2 \delta^2 \ell_n^2}{(1+\delta)^2} (1+\delta)n/\ell_n} + e^{-\frac{2 \delta^2 \ell_n^2}{(1-\delta)^2} (1-\delta)n/\ell_n} \nonumber \\
    &\leq 2e^{-\frac{2 \delta^2 \ell_n}{(1+\delta)} n}
 \end{align}
  Define $V_i \triangleq \mathbf{1}_{\left\{N_i =\lceil r\log{n}\rceil,B\right\}}$ and $\Tilde{V} = \sum_{i=1}^{n/\ell_n}V_i$. It is clear that $||\Y_r| - \Tilde{V}| \leq |K - n/\ell_n|$. Moreover $E[\Tilde{V}] = \frac{n}{\ell_n}q_{r,n}e_{r,n}$. Therefore
\begin{align}
    \text{Pr}&\left(\middle||\Y_r| - nq_{r,n}e_{r,n}/\ell_n\middle| \geq \epsilon_{r,n}n/\ell_n\right) \nonumber \\
    \stackrel{(a)}{\leq} \text{Pr}&\left(\middle|\Tilde{Z} - nq_{r,n}e_{r,n}/\ell_n\middle| \geq n\epsilon_{r,n}/(2\ell_n)\right) \nonumber + \text{Pr}\left(|K - n/\ell_n| \geq n\epsilon_{r,n}/(2\ell_n)\right) \nonumber \\
    &\leq 2e^{-n\epsilon_{r,n}^2/(2\ell_n)} + 2e^{-\frac{8 \epsilon_{r,n}^2 \ell_n}{(1+2\epsilon_{r,n})} n},
\end{align}
where (a) is due to triangle inequality. We then consequently use Hoeffding's Inequality and (\ref{eq:BoundK}).
\newpage
\section{Proof of Lemma~\ref{lemma:BigPiecesNo}}\label{app:BigPiecesNo}
\BigPiecesNo*
Consider Lemma~\ref{Lemma:ConcBoundCoverageNoisy}. Since $\gamma$ is a value we can choose, choose it to be $ (J/L)$.
Note that
\begin{align*}
    F_{d}\{\gamma\log{n}\} &=\frac{E\left[N_1\mathbf{1}_{\{\vec{X}_i \in \Y_{\geq g}\}}\right]}{\ell_n}.
\end{align*}
Now define the event 
\begin{align*}
\mathcal{J} \triangleq \left\{\Co_{d}\{\gamma\log{n}\} > E\left[N_1\mathbf{1}_{\{\vec{X}_i \in \Y_{\geq g}\}}\middle/\ell_n\right] + \delta\right\}.
\end{align*}
From the general version of Lemma~\ref{Lemma:ConcBoundCoverage}, $\Pr(\mathcal{J}) \to 0$ as $n \to \infty$. Therefore we can write
\begin{align*}
    &H(\Y_{\geq J}) \\
    &\leq H(\Y_{\geq J}|\overline{\mathcal{J}}) + H(\Y_{\geq J}|\mathcal{J})\Pr(\mathcal{J}) + 1 \\
    &\leq H(\Y_{\geq J}|\overline{\mathcal{J}}) + 2n\Pr(\mathcal{J}) + 1 \\
    &\leq 2n\left(E\left[N_1\mathbf{1}_{\{\vec{X}_i \in \Y_{\geq g}\}}\middle]\right/\ell_n + \delta\right) + o(n) \\
    &\stackrel{(a)}{\leq} 2n\left(E\left[N_1\mathbf{1}_{\{N_1 \geq (J/L)\log{n}\}}\middle/\ell_n\right] + \delta\right) + o(n) \\
    &\stackrel{(b)}{\leq} 2n\left(\sqrt{E[N_1^2]\Pr(N_1 \geq (J/L)\log{n})}/\ell_n + \delta\right) + o(n) \\ 
    &\stackrel{(c)}{\leq} 2n\left(\sqrt{E[N_1^2]\ell_nL}/(\ell_n\sqrt{J\log{n}})\right) + \delta) + o(n) \\
    &\stackrel{(d)}{\leq} 2n\left(S\sqrt{L/J} + \delta\right) + o(n),
\end{align*}
for some finite $S$. $(a)$ is true because we include the deleted fragments. $(b)$ is due to Cauchy-Schwarz inequality. $(c)$ is due to Markov Inequality and $(d)$ is due to $E[N_1/\log{n}]$ and $E[(N_1/\log{n})^2]$ being finite and bounded. Therefore this implies
\begin{align*}
    \lim_{n \to \infty}\frac{H(\Y_{\geq J})}{n} \leq 2\left(S\sqrt{L/J} + \delta\right).
\end{align*}
\newpage

\section{Proof of Lemma~\ref{lemma:DistractionTerms}}\label{app:DistractionTerms}
\DistractionTerms*
This lemma can proved by careful splitting of terms and careful manipulation of upper bounds.
First we have
\begin{align}\label{eq:subordertermsconvergence}
    &\lim_{n \to \infty} A(n) 
    = \lim_{n \to \infty}\sum_{k=1}^{J}\left(\frac{1}{n} + 2\text{Pr}(\mathcal{E}_{k,n}) + \frac{M_k}{n}(\log{e} + P)\right) + \nonumber \\
    &\lim_{n \to \infty}\epsilon_n\sum_{k=1}^{J}\left(\left(\frac{k}{L}-1\right)^+ + \frac{(\epsilon_n + 1)\log{(\ell_n\log{n})\log{(n)}}}{\ell_n}\right) \nonumber \\
    &\stackrel{(a)}{\leq} \lim_{n \to \infty} \sum_{k=1}^{J}\left(\frac{1}{n} +  2e^{-n\epsilon^2/(2\ell_n)} + 2e^{-\frac{8 \epsilon^2 \ell_n}{(1+2\epsilon)} n}\right) 
    \nonumber \\
    &+\lim_{n \to \infty} \sum_{k=1}^{J}\frac{M_k}{n}(\log{e} + P) \nonumber \\
    &= J\lim_{n \to \infty} \left(\frac{1}{n} +  2e^{-n\epsilon^2/(2\ell_n)} + 2e^{-\frac{8 \epsilon^2 \ell_n}{(1+2\epsilon)} n}\right) \nonumber \\
    &+ \lim_{n \to \infty} \sum_{k=1}^{\infty}\frac{M_k}{n}(\log{e} + P) =\lim_{n \to \infty} \sum_{k=1}^{\infty}\frac{M_k}{n}(\log{e} + P),
\end{align}
where (a) is due to lemma~\ref{lemma:y_kconc} and the fact that we set $\epsilon_n = 1/\log{n}$.
. We proceed as
\begin{align}
     &\sum_{k=1}^{J}\frac{M_k}{n}(\log{e} + P) = \frac{1}{\ell_n}\log{e}\sum_{k = 1}^{J}(q_{k,n}e_{k,n} + \epsilon_n) + \sum_{k = 1}^{J}\frac{M_k}{n}P \nonumber \\
    &\leq \frac{\log{e}}{\ell_n}\left(1 + J\epsilon_n\right) + \sum_{k=1}^{L-1}\frac{M_k}{n}P + \sum_{k=L}^{J}\frac{M_k}{n}P \nonumber \\
    &\leq o(1) + \sum_{k=1}^{L-1}\frac{M_k}{n}\log\left(1 + \frac{2n^{k/L}-1}{M_k}\right) \nonumber \\
    &+ \sum_{k=L}^{J}\frac{M_k}{n}\log\left(2 + \frac{M_k-1}{n^{k/L}}\right)
\end{align}
since $\min(a,b) \leq a,b$. 
Let's now look at the second term
\begin{align}\label{eq:SecondReal}
    \sum_{k=1}^{L-1}\frac{M_k}{n}\log\left(1 + \frac{2n^{k/L}-1}{M_k}\right) 
    &\leq \sum_{k=1}^{L-1} \frac{M_k}{n}\frac{2n^{k/L}}{M_k}  \nonumber \\
    = 2\sum_{k=1}^{L-1}n^{k/L - 1} &\to 0,
\end{align}
as $n \to \infty$. For a finite $L$, this is a finite summation. Since for $k < L$, $n^{k/L - 1} \to 0$ as $n \to \infty$, the above summation also goes to zero. Now look at the third term
\begin{align}\label{eq:ThirdReal}
    &\sum_{k=L}^{J}\frac{M}{n}\log\left(2 + \frac{M_k-1}{n^{k/L}}\right) \leq \frac{1}{n}\sum_{k=L}^{J}M_k(1 + (M_k-1)n^{-k/L}) \nonumber \\
    &\leq \frac{1}{\ell_n}\sum_{k=L}^{J}(q_{k,n}e_{k,n} + \epsilon_n) + \frac{1}{\ell_n^2}\sum_{k=L}^{J}n^{1 - k/L}(q_{k,n}e_{k,n} + \epsilon_n)^2 \nonumber \\
    &\leq \frac{1}{\ell_n}E[\mathbf{1}_{B'}] + \frac{1}{\ell_n^2}\sum_{k=L}^{\infty}n^{1-k/L} + 3J\epsilon_n/\ell_n + \epsilon_n^2/\ell_n\nonumber \\
    &\leq \frac{1}{\ell_n} + \frac{1}{(1 - n^{-1/L})\ell_n^2} + o(1),
\end{align}
which $\to 0$ as $n \to \infty$. Equations \eqref{eq:subordertermsconvergence}, \eqref{eq:SecondReal} and \eqref{eq:ThirdReal} imply that $A(n) \to 0$ as $n \to \infty$.
\newpage
\section{Proof of Lemma~\ref{lem:NoisyUpperBound}}\label{app:NoisyUpperBound}
\NoisyUpperBound*

Consider the noisy shuffling channel from \cite{noisyshuffling}. Let $\vec X_r$ be the input binary string, which consists of $M_r$ strings of length (exactly) $r\log{n}$ in order (thus, this can also be viewed as a single string of length $M_r(r\log{n})$). Let $\vec Y_r$ be the output binary string. The first step towards proving Lemma~\ref{lem:NoisyUpperBound}, is to establish a relationship between the entropies $H(\vec Y_r)$ and  $H(S^{M_r}|\vec X_r,\vec Y_r)$, where $S^{M_r}$ is a shuffling vector, formally defined in the next paragraph.

We define 
$$\vec X_r := \left[\tilde X_1^{r\log{n}}, \tilde X_2^{r\log{n}},\dots,\tilde X_{M_r}^{r\log{n}}\right],$$
where $\tilde X_i^{r\log{n}}$ is the $i$th string of length $r\log{n}$. Let 
$$\vec Z_r = \left[\tilde Z_1^{r\log{n}}, \tilde Z_2^{r\log{n}},\dots,\tilde Z_{M_r}^{r\log{n}}\right],$$
be a random error pattern created by passing $\vec X_r$ through a BSC($p$). Further, given $M_r$, define $S^{M_r} \in [1:M_r]^{M_r},$ as a shuffle vector, uniformly distributed on all possible permutations of the indices $[1:M_r]$. Now, similar to $\vec X_r$, we can define
$$\vec Y_r := \left[\tilde Y_1^{r\log{n}}, \tilde Y_2^{r\log{n}},\dots,\tilde Y_{M_r}^{r\log{n}}\right],$$
where $\tilde Y_i^{r\log{n}}:= X_{S(i)}^{r\log{n}}\oplus Z_{S(i)}^{r\log{n}}.$ The following lemma adapted from \cite{noisyshuffling} is central to proving Lemma~\ref{lem:NoisyUpperBound}.



\begin{lemma}\label{lem:NoisyShuffling}
Let $\vec Y_r$ 
be obtained by passing string $\vec{X}_r$ through the noisy shuffling channel. Let $S^{M_r} \in \{1:M_r\}^{M_r}$ be the uniform shuffling vector, encoding the random shuffle induced by this channel. For this channel, if

\begin{align}
    1 - H(2p) -\frac2r \geq 0,
\end{align}
then
\begin{align}\label{eq:boundtight}
    H(\vec Y_r) + H(S^{M_r}|\vec X_r,\vec Y_r)) \leq M_r(\lceil r\log{n}\rceil) + M_r^{1-\delta^{\prime}}\log{M_r},
\end{align}  
for some \(\delta^{\prime} > 0\). Otherwise,
\begin{align}\label{eq:boundloose}
     H(\vec Y_r) + H(S^{M_r}|\vec X_r,\vec Y_r)) \leq M_r(\lceil r\log{n}\rceil) + M_r\log{M_r}.
\end{align}
\end{lemma}



The proof of the above lemma follows from very similar arguments as \cite{noisyshuffling}. Equation \eqref{eq:boundtight} is similar to Lemma~1 in \cite{noisyshuffling}, with a key difference.  Lemma~1 in \cite{noisyshuffling} uses the following upper bound:
\begin{align}
    H(\vec Y_r) + H(S^{M_r}|\vec X_r,\vec Y_r)) \leq M_r(\lceil r\log{n}\rceil) + o(M_rr\log{n}).
\end{align}

During the proof of Lemma~1 in \cite{noisyshuffling}, a loose upper bound is used to bound the term \(M_r^{1-\delta}\log{M_r} \leq M_r r\log{n}\), where \(\delta > 0\). Instead, we retain \(M_r^{1-\delta}\log{M_r}\) in Lemma~\ref{lem:NoisyShuffling}.

Equation~\eqref{eq:boundloose} can be shown as follows:
\begin{align}
     H(\vec Y_r) + H(S^{M_r}|\vec X_r,\vec Y_r)) \stackrel{(a)}\leq H(\vec Y_r) + H(S^{M_r}) \stackrel{(b)}\leq M_r(\lceil r\log{n}\rceil) + M_r\log{M_r},
\end{align}

where \((a)\) follows because conditioning reduces entropy and \((b)\) is because the uniform distribution maximizes entropy.


To utilize this lemma, define $\Y_r$ as an unordered multiset which contains $M_r$ pieces of size $r\log{n}$. 
We can view the output of the noisy shuffling channel as either $\Y_r$ or $\vec Y_r$, since the shuffling channel encodes a uniformly random shuffling on the input strings. 
Further we assume $M_r$ to be a random variable, instead of a fixed quantity. 
Precisely $(M_r:r\in A)$ are random variables that have a joint conditional distribution of $(|\Y_r| : r \in A)$ conditioned on $\bar{\mathcal{E}}_{k,n}$. We assume that the random variable \(M_r\) is implicitly conditioned on event \(\bar{\mathcal{E}}_{k,n}\) for the rest of this proof.
We can now say that
$$I(\vec{X}_r;\Y_r) = I(\vec{X}_r;\vec Y_r).$$
The left hand side of the above equation can be further expanded as 
\begin{align}\label{eq:LHSChannelEquivalence}
    &I(\vec{X}_r;\Y_r) = H(\Y_r) - H(\Y_r|\vec X_r) = H(\Y_r) + H(\vec Z_r|\vec X_r,\Y_r) - H(\vec Z_r) -H(\Y_r|\vec X_r, \vec Z_r) \nonumber \\
    &\stackrel{(a)}{=}  H(\Y_r) + H(\vec Z_r|\vec X_r, \Y_r) - H(\vec Z_r),
\end{align}
where $(a)$ follows, since given $(\vec X_r, \vec Z_r)$, we can calculate $\Y_r$.
Lemma~\ref{lem:NoisyShuffling} applies when we condition each entropy term in the bounds \eqref{eq:boundtight} and \eqref{eq:boundloose} as $M_r = m_r$ for any fixed number $m_r$.
Now following the steps in equations~(7) and (8) in \cite{noisyshuffling} we can argue (for a fixed length channel with output containing pieces of size $r\log{n}$ and cardinality $M_r$),
\begin{align}\label{eq:RHSChannelEquivalence}
    I(\vec{X}_r;\vec Y_r) &= H(\vec Y_r) + H(S^{M_r}|\vec{X}_r,\vec Y_r) - H(S^{M_r},\vec Z_r,\vec Y_r|\vec X_r) \nonumber \\
    &= H(\vec Y_r) + H(S^{M_r}|\vec{X}_r,\vec Y_r) - H(S^{M_r}) - H(\vec{Z}_r).
\end{align}
Therefore from \eqref{eq:LHSChannelEquivalence} and \eqref{eq:RHSChannelEquivalence}, we conclude that
\begin{align}\label{eq:fullupper}
    H(\Y_r) &+ H(\vec Z_r|\vec X_r, \Y_r) =  H(\vec Y_r) + H(S^{M_r}|\vec{X}_r,\vec Y_r) -H(S^{M_r}) \nonumber \\
    &\stackrel{(a)}{\leq} H(\vec Y_r,M_r) + H(S^{M_r}|\vec{X}_r,\vec Y_r,M_r) -H(S^{M_r}|M_r) \nonumber \\
    & \leq H(M_r) + H(\vec Y_r|M_r) + H(S^{M_r}|\vec{X}_r,\vec Y_r,M_r) -H(S^{M_r}|M_r) \nonumber \\
    &\stackrel{(b)}{\leq} \log{(n+1)} + H(\vec Y_r|M_r) + H(S^{M_r}|\vec{X}_r,\vec Y_r,M_r) -H(S^{M_r}|M_r),
\end{align}
where $M_r$ is viewed as a random variable. Step $(a)$ follows because $M_r$ is a deterministic function of $\vec Y_r$ and conditioning reduces entropy, and $(b)$ follows because there are at most $n+1$ values that $M_r$ can take. To see this note that \(M_r\) is defined as the number of elements in \(|\Y_r|\), and this cannot exceed the length of the length of the input string \(n\).

Now to prove Lemma~\ref{lem:NoisyUpperBound}, define $\mathcal{A} := \mathbb{Z}^+\cap \left[\frac{k-1}{L}\log{n},\frac{k}{L}\log{n}\right)$. 
We can use Lemma~\ref{lem:NoisyShuffling}, when conditioned on $M_r = m_r$, where $m_r$ is a fixed number. The following upper bound then follows:
\begin{align}\label{eq:finalub}
    &H(\Y_k|\bar{\mathcal{E}}_{k,n}) + H(\vec \Z_k|\vec \X_k,\Y_k,\bar{\mathcal{E}}_{k,n}) \leq \sum_{r: r\log{n}\in \mathcal{A}}\left(H(\Y_r|\bar{\mathcal{E}}_{k,n}) + H(\vec Z_r|\vec X_r,\Y_r,\bar{\mathcal{E}}_{k,n})\right) \nonumber \\
    &\stackrel{(a)}{\leq} \sum_{r: r\log{n}\in \mathcal{A}}\left(\log{(n+1)} + H(\vec Y_r|M_r) + H(S^{M_r}|\vec{X}_r,\vec Y_r,M_r) -H(S^{M_r}|M_r)\right) \nonumber \\
    &\stackrel{(b)}{\leq} \frac{\log^2{(n+1)}}{L} + \sum_{r: r\log{n}\in \mathcal{A}}\sum_{m_r\in[0:n]}\Pr(m_r)\left( H(\vec Y_r|m_r) + H(S^{M_r}|\vec{X}_r,\vec Y_r,m_r) -H(S^{M_r}|m_r)\right)\nonumber \\
    &\stackrel{(c)}{=} \frac{\log^2{(n+1)}}{L} + \sum_{r: r\log{n}\in \mathcal{A}}\sum_{m_r\in[0:n]}\Pr(m_r)\left( H(\vec Y_r|m_r) + H(S^{M_r}|\vec{X}_r,\vec Y_r,m_r) -\log{(m_r!)}\right),
\end{align}
where $(a)$ is due to \eqref{eq:fullupper}
$(b)$ is because $|\mathcal{A}|\leq \log{n}/L$ and $(c)$ is because $S^{M_r}$ given $\{M_r = m_r\}$ is a uniform shuffling vector, which picks one of exactly $m_r!$ values, equiprobably. Note that we use the shorthand $m_r$ to refer to the event $\{M_r = m_r\}$. Recall that \(M_{k} := \left(\epsilon_{n} +q_{k,n}\right)\frac{n}{\ell_n}\), which is a constant. 

 Firstly,
when $\frac{k-1}{L} > \frac{2}{1-H(2p)}$ we can upper bound \eqref{eq:finalub} as
\begin{align}\label{eq:jensen}
    &H(\Y_k|\bar{\mathcal{E}}_{k,n}) + H(\vec Z_k|\vec X_k,\Y_k,\bar{\mathcal{E}}_{k,n}) \nonumber \\
    &\stackrel{(a)}{\leq} \frac{\log^2{(n+1)}}{L} + \sum_{r: r\log{n}\in \mathcal{A}}E\left(M_r(\lceil r\log{n}\rceil) + M_r^{1-\delta^{\prime}}\log{M_r} - \log{(M_r!)}\right) \nonumber \\
    &\stackrel{(b)}{\leq} \frac{\log^2{(n+1)}}{L} + \sum_{r: r\log{n}\in \mathcal{A}}E[M_r]\frac kL\log{n} + \sum_{r: r\log{n}\in \mathcal{A}}M_r^{1-\delta^{\prime}}\log{M_k} - \sum_{r: r\log{n}\in \mathcal{A}}\log{(M_r!)}   \nonumber \\
    &\stackrel{(c)}\leq \frac{\log^2{(n+1)}}{L} + M_k\frac{k}{L}\log{n} + o(M_k\log{M_k}) - \sum_{r: r\log{n}\in \mathcal{A}}E[M_r\log(M_r)] \nonumber \\
    &\stackrel{(d)}{\leq}  \frac{\log^2{(n+1)}}{L} + M_k\frac{k}{L}\log{n} + o(M_k\log{M_k}) - \sum_{r: r\log{n}\in \mathcal{A}}E[M_r]\log(E[M_r])\nonumber \\
    &\stackrel{(e)}{\leq}  \frac{\log^2{(n+1)}}{L} + M_k\frac{k}{L}\log{n} + o(M_k\log{M_k}) - \left(M_k-\frac{2n\epsilon_n}{\ell_n}\right)\log\left(M_k - \frac{2n\epsilon_n}{\ell_n}\right),
\end{align}
where $(a)$ is due to the first part of Lemma~\ref{lem:NoisyShuffling}, $(b)$ is because of linearity of expectation followed by an upper bound on the sum of $m_r \in \mathcal{A}$, upper bounding each term \(\log{M_r}\) with \(\log{M_k}\) and Stirling's approximation, 
and $(c)$ is due to Jensen's inequality for the convex function $x^{1-\delta^{\prime}}$. 
To obtain $(d)$, we upper bound the last term in \eqref{eq:jensen} as
\begin{align}
    &-\sum_{r: r\log{n}\in \mathcal{A}}E[M_r]\log(E[M_r]) \stackrel{(i)}{\leq}  -\sum_{r: r\log{n}\in \mathcal{A}}(E[M_r])\times \log{\left(\frac{\sum_{r: r\log{n}\in \mathcal{A}}E[M_r]}{\sum_{r: r\log{n}\in \mathcal{A}} 1}\right)} \nonumber \\
    &\stackrel{(ii)}{\leq}  -\left(M_k-\frac{2n\epsilon_n}{\ell_n}\right)\log\left(M_k - \frac{2n\epsilon_n}{\ell_n}\right) + \left(M_k-\frac{2n\epsilon_n}{\ell_n}\right)\log{\left(\frac{\log{n}}{L}\right)} \nonumber \\
    &\stackrel{(iii)}{=} -\left(M_k-\frac{2n\epsilon_n}{\ell_n}\right)\log\left(M_k - \frac{2n\epsilon_n}{\ell_n}\right) + o(M_k\log{M_k}),
\end{align}
where $(i)$ is due to log-sum inequality, $(ii)$ is because $\sum_{r: r\log{n}\in \mathcal{A}}E[M_r] = E[|\Y_k|] \geq \left(M_k-2n\epsilon_n/\ell_n\right)$ (see \eqref{eq:upperlowercard}) and \((iii)\) is because from the definition of \(M_{k} := \left(\epsilon_{n} +q_{k,n}\right)\frac{n}{\ell_n}\) one can see that \(M_k \log n = o(M_k \log M_k)\).


On the contrary if \(\frac{k-1}{L} \leq \frac{2}{1-H(2p)}\)
we opt for the looser upper bound to show that
\begin{align}
&H(\Y_k|\bar{\mathcal{E}}_{k,n}) + H(\vec Z_k|\vec X_k,\Y_k,\bar{\mathcal{E}}_{k,n}) \nonumber \\
    &\stackrel{(a)}{\leq} \frac{\log^2{(n+1)}}{L} + \sum_{r: r\log{n}\in \mathcal{A}}E\left(M_r(\lceil r\log{n}\rceil) + M_r\log{M_r} - \log{(M_r!)}\right) \nonumber \\
    &\leq  \frac{\log^2{(n+1)}}{L} + M_k\frac{k}{L}\log{n} + o(M_k\log{M_k})
\end{align}
This proves the theorem.
\newpage
\section{Proof of Lemma~\ref{lem:NoisyTermsConsolidation}}\label{app:NoisyTermsConsolidation}
\NoisyTerms*
We apply Lemma~\ref{lem:NoisyUpperBound}, followed by careful analysis to prove Lemma~\ref{lem:NoisyTermsConsolidation}. Steps are similar to \eqref{eq:SetUpperBound}, \eqref{eqn:FanoFirstBound} and \eqref{eq:deltabound}, with some minor changes. Precisely
\begin{align}    &\sum_{k=L+1}^{J}\left(H(\Y_k|\bar{\mathcal{E}}_{k,n}) + H(\vec{\Z}_{k}|\vec{\X}_{k},\Y_{k},\bar{\mathcal{E}}_{k,n})\right) \nonumber \\
    &\stackrel{(a)}{\leq} \sum_{k=L+1}^{\frac{2L}{1-H(2p)}} \left(M_k\frac{k}{L}\log{n}\right) + \sum_{k = \frac{2L}{1-H(2p)} + 1}^{J} \left(M_k\frac{k}{L}\log{n} - \left(M_k-\frac{2n\epsilon_n}{\ell_n}\right) \log{\left(M_k-\frac{2n\epsilon_n}{\ell_n}\right)}\right) \nonumber \\
    &+ \sum_{k=L+1}^{J} o(M_k\log{M_k}) + \frac{J\log^2{(n+1)}}{L} \nonumber \\
    &\stackrel{(b)}{\leq}  \sum_{k=L+1}^{\frac{2L}{1-H(2p)}}\left(\frac{n}{\ell_n}(\epsilon_n + q_{k,n})\right) \frac kL\log{n}  \nonumber \\ 
    &+ \sum_{k = \frac{2L}{1-H(2p)} + 1}^{J}\left(\left(\frac{n}{\ell_n}(\epsilon_n + q_{k,n})\right) + \frac kL\log{n}\left(\frac{n}{\ell_n}(-\epsilon_n + q_{k,n})\right)\log{\left(\frac{n}{\ell_n}(-\epsilon_n + q_{k,n})\right)}\right) \nonumber \\
    &+  \sum_{k=L+1}^{J} o(M_k\log{M_k}) + \frac{J\log^2{(n+1)}}{L}\nonumber \\
    &\stackrel{(c)}{\leq} \frac{n}{\ell_n}\sum_{k= L+1}^{\infty}q_{k,n}\frac{k}{L}\log{n} - \frac{n\log{n}}{\ell_n}\sum_{k = \frac{2L}{1-H(2p)}}^{\infty}q_{k,n} + o(n) \nonumber \\
    &= n\left(\frac{E\left[N_1\mathbf{1}_{\{N_1\geq\log{n}\}}\right]}{\ell_n} - \frac{\log{n}}{\ell_n}E\left[\mathbf{1}_{\left\{N_1\geq\frac{2}{1-2H(p)}\log{n}\right\}}\right]\right) + o(n),
\end{align}
where $(a)$ is due to Lemma~\ref{lem:NoisyUpperBound}, $(b)$ is from the definition of $M_{k} = \frac{n}{\ell_n}(q_{k,n} + \epsilon_{n})$, $(c)$ is by collecting the ancillary terms and grouping them. These terms can be shown to be $o(n)$ using arguments similar to the ones used in \eqref{eq:subordertermsconvergence}.
\section{Proof of Lemma~\ref{lem:noisydistraction}}\label{app:noisydistraction}
\NoisyDistraction*
This lemma again involves careful manipulation of upper bounds, to arrive at the result.
\begin{align}\label{eq:NoisyDistraction}
    &\frac{1}{n}\sum_{k=1}^{L}H(\Y_{k}|\bar{\mathcal{E}}_{k,n}) +  2\sum_{k=1}^{J}\left(2\Pr(\mathcal{E}_{k,n}) + \frac{1}{n}\right) \nonumber \\
    &\stackrel{(a)}{\leq} \sum_{k=1}^L \left(\frac{M_k\log{n}}{n}\left(\frac{k}{L} - 1\right)^+ + \frac{M_k}{n}\left(\log{e} + P\right)+ \frac{M_k\log{n}\log{(\ell_n\log{n})}}{n}\right) \nonumber \\
    &+  2\sum_{k=1}^{J}\left(2\Pr(\mathcal{E}_{k,n}) + \frac{1}{n}\right) \stackrel{(b)}{\leq} \sum_{k=1}^L \left(\frac{M_k}{n}\left(\log{e} + P\right)+ \frac{M_k\log{n}\log{(\ell_n\log{n})}}{n}\right) \nonumber \\
    &+ 2J\left(\frac1n + 2e^{-n\epsilon_n^2/(2\ell_n)} + 2e^{-\frac{8 \epsilon_n^2 \ell_n}{(1+2\epsilon_n)} n}\right)
    \stackrel{(c)}{\leq} \sum_{k=1}^{L-1}\frac{M_k}{n}\left(\log{e} + \log{ \left(1 + \frac{2n^{k/L}-1}{M_k}\right)} \right)\nonumber \\
    &+ \frac{M_L}{n}\log{\left(e\left(1+\frac{2n-1}{M_L}\right)\right)} + 2J\left(\frac1n + 2e^{-n\epsilon_n^2/(2\ell_n)} + 2e^{-\frac{8 \epsilon_n^2 \ell_n}{(1+2\epsilon_n)} n}\right) \nonumber \\
    &\leq \sum_{k=1}^{L-1} \frac{M_k}{n}\left(\log{e} + \frac{2n^{k/L}}{M_k}\right) + \frac{\log\left(e\left(1+\frac{\ell_n}{\epsilon_n}\right)\right)}{\ell_n} + 2J\left(\frac1n + 2e^{-n\epsilon_n^2/(2\ell_n)} + 2e^{-\frac{8 \epsilon_n^2 \ell_n}{(1+2\epsilon_n)} n}\right) \nonumber \\
    &\leq 2\sum_{k=1}^{L-1}n^{k/L - 1} +  L(1+\epsilon_n)\frac{\log{e}}{\ell_n} + O\left(\frac{\log\left(1+\log^2{n}\right)}{\log{n}}\right)\nonumber \\
    &+ 2J\left(\frac1n + 2e^{-n\epsilon_n^2/(2\ell_n)} + 2e^{-\frac{8 \epsilon_n^2 \ell_n}{(1+2\epsilon_n)} n}\right)\to 0,
\end{align}
    as $n \to \infty$,
where recall that 
\begin{align}
    P := \min\left(\log\middle(2 + \frac{M_k-1}{n^{k/L}}\middle),
    \log\middle(1 + \frac{2n^{k/L}-1}{M_k}\middle)\right).
\end{align}
\newpage


\end{document}